\documentclass[12pt]{article}
\usepackage{amsfonts}
\usepackage{latexsym,amsmath}
\usepackage{amssymb,array}
\makeatletter \oddsidemargin 0in \evensidemargin 0in \textwidth
16cm\textheight 20cm 
\newcommand{\ov}{\overline}
\newcommand{\wt}{\widetilde}
\newtheorem{t1}{Theorem}[section]
\newtheorem{d1}{Definition}[section]

\newtheorem{l1}{Lemma}[section]
\newtheorem{r1}{Remark}[section]
\newtheorem{p1}{Proposition}[section]

\begin{document}
\title{ON SOME GENERALIZED AGEING ORDERINGS}
\author{Asok K. Nanda\footnote{e-mail: asok.k.nanda@gmail.com, corresponding author.},
 Nil Kamal Hazra 
\\Department of Mathematics and Statistics
 \\IISER Kolkata, Mohanpur Campus
\\Mohanpur 741252, India\\
\and\\
\\D.K. Al-Mutairi, M.E. Ghitany
\\Department of Statistics and Operations Research
\\Faculty of Science, Kuwait University
\\P.O. Box 5969 Safat 13060, Kuwait}
\date{August, 2014}
\maketitle
\begin{abstract}
Some partial orderings which compare probability distributions with the exponential distribution, are found to be very useful to understand the phenomenon of ageing. 
Here, we introduce some new generalized partial orderings which describe the same kind of characterization of some generalized ageing classes. We give some equivalent conditions for each of the orderings. 
Inter-relations among the generalized orderings
have also been discussed.

\end{abstract}
{\bf Key words:} Generalized ageing classes, Lorenz curve, partial ordering, TTT transform.
\section{Introduction}\label{sec1}
\hspace*{0.2 in}Ageing and partial ordering are two very well known concepts in reliability theory. Positive ageing describes the situation where an older system has shorter remaining lifetime
in some stochastic sense than a younger one. Many classes of lifetime distributions are characterized by their ageing properties. Exponential distribution is exceptional one
which has no ageing property due to its memory less property. It will not be out of the way to mention here that there are some orderings for which Weibull distribution 
is the borderline distribution, namely, ageing intensity ordering, see, for example, Nanda et al.~\cite{nba1}, Righter et al.~\cite{rss1}.
 Many different types of ageing notions have been studied in the literature, for instance, see Bryson and Siddiqui~\cite{bs6}, Barlow and Proschan~\cite{bp1}, 
 Klefsj$\ddot{\rm{o}}$~\cite{kle1}, Deshpande et al.~(\cite{dks1}, \cite{dsbj1}),
 Loh~\cite{lw1}, Lai and Xie~\cite{lx1} and the references there in.
 On the other hand, partial orderings are used to compare two different distributions. Shaked and Shanthikumar~\cite{shak1} is a very good reference for this purpose. 
 It has been observed that among all the partial orderings, there are two special kinds of partial orderings which describe the phenomenon of ageing: Firstly, the partial orderings which compare 
 probability distributions with the exponential distribution; secondly, those which compare residual lifetimes at different ages. In our paper we concentrate our discussion particularly 
 on the first case. 
 The significant works in the direction of our work have been developed by Kochar and Wiens~\cite{kw1}, Sengupta and Deshpande~\cite{sd6} and many other researchers.
%%%%%%%%%%%%%%%%%%%%%%%%%%%%%%%%%%%%%%%%%%%%%%%%%%%%%%%%%%%%%%%%%%%%%%%%%%%%%%%%%%%%%%%%%%%%%%%%%%%%%%%%%%%%%%%%%%%%
\\\hspace*{0.3 in}For an absolutely continuous nonnegative random variable $X$, the probability density function is denoted
 by $f_X(\cdot)$ and the distribution function by $F_X(\cdot)$. We write $\bar F_X(\cdot)\equiv 1-F_X(\cdot)$ to denote the 
survival function of the random variable $X$.
%\\\hspace*{0.2 in}For any absolutely continuous nonnegative random variable $X$ with probability density function 
%$f(\cdot)$ and survival function $\overline{F}(\cdot)$, 
Let us write
$$\overline{T}_{X,0}(x)=f_X(x),$$
and
\begin{equation}
\overline{T}_{X,s}(x)=\frac{\int_x^\infty \overline{T}_{X,s-1}(t)dt}{\wt\mu_{X,s-1}},\label{eqn1-1}
\end{equation}
for $s=1,2,\ldots$, where
$$\wt\mu_{X,s}=\int_0^\infty \overline{T}_{X,s}(t)dt,$$
$s=0,1,2,\ldots$, $\cdot$ We assume $\wt \mu_{X,s}$ to be finite. Note that $\overline{T}_{X,2}(\cdot)$
is the survival function of the equilibrium distribution of $X$, which plays an important
role in ageing concepts (Deshpande et al.~\cite{dks1}), whereas $\overline{T}_{X,s}(\cdot)$ is the 
survival function of the equilibrium distribution of a distribution with survival 
function $\overline{T}_{X,s-1}(\cdot)$, $s=1,2,\ldots\cdot$ We further define, for $s=1,2,\dots,$
\begin{eqnarray*}
r_{X,s}(x)&=&\frac{\overline{T}_{X,s-1}(x)}{\int_x^\infty \overline{T}_{X,s-1}(t)dt}\\
&=&\frac{\overline{T}_{X,s-1}(x)}{\wt \mu_{X,s-1}\overline{T}_{X,s}(x)},
\end{eqnarray*}
and 
\begin{eqnarray*}
 \mu_{X,s}(x)&=&\frac{\int_x^\infty \overline{T}_{X,s}(t)dt}{\overline{T}_{X,s}(x)},
\end{eqnarray*}

where $r_{X_s}(\cdot)$ and $\mu_{X,s}(\cdot)$, respectively, represent the failure rate and the mean residual life functions corresponding to $\overline{T}_{X,s}(\cdot)$.
%For $s=1$, $r_{X,1}(x)$ is the failure rate of $X$, defined as the ratio of
%the probability density to the survival, whereas, for $s=2$, $r_{X,2}(t)$ is the reciprocal
%of the mean residual life $E(X_t)=E\left(X-t|X\ge t\right)$. 
Note that, for $s=1,2,\dots,$
$$\mu_{X,s}(0)=\wt \mu_{X,s},$$ and, for $s=2,3,\dots,$
\begin{eqnarray}\label{mre5}
r_{X,s}(x)=\frac{1}{\mu_{X,s-1}(x)}.
\end{eqnarray}
Let ${\cal F}$ be the class of distribution functions $F:[0,\infty)\longrightarrow [0,1]$
with $F(0)=0$. We assume that all $F(\in {\cal F})$ have their finite generalized means $\wt \mu_{X,s}$, and are strictly
increasing on their support. If $F$ is not strictly increasing, we take
the inverse as 
$$F^{-1}(y)=\inf\{x:F(x)\ge y\}.$$
Throughout the paper, increasing and decreasing properties are not used in strict sense. For any differentiable function $k(\cdot)$, we write $k'(t)$ to denote the first 
derivative of $k(t)$ with respect to $t$.
%%%%%%%%%%%%%%%%%%%%%%%%%%%%%%%%%%%%%%%%%%%%%%%%
% We write, for $x\geq 0$ and $s=1,2,\dots,$
% $$\Phi_{X,s}(x)=\frac{1}{\tilde \mu_{X,s}}\int\limits_0^{x}\ov T_{X,s}(y)dy$$
% and $$\ov \Phi_{X,s}(x)=1-\Phi_{X,s}(x)=\frac{1}{\tilde \mu_{X,s}}\int\limits_x^{\infty}\ov T_{X,s}(y)dy.$$
\\\hspace*{0.2 in}The scaled total time on test (TTT) transform is a very useful tool to analyze the statistical lifetime data. It was first introduced by Barlow and Campo~\cite{bc1}. 
To know more about TTT transform,
readers may refer to Barlow~\cite{b6} and the references there in.
The TTT transform corresponding to $\overline{T}_{X,s}(\cdot)$ is denoted by $\mathcal {H}^{-1}_{X,s}(\cdot)$, and is defined as
\begin{eqnarray*}
\mathcal {H}^{-1}_{X,s}(u)&=&\frac{1}{\wt \mu_{X,s}}\int\limits_0^{T^{-1}_{X,s}(u)}\ov T_{X,s}(y)dy
\\&=&T_{X,s+1}\left(T^{-1}_{X,s}(u)\right),
\end{eqnarray*}
%%%%%%%%%%%%%%%%%%%%%%%%%%%%%%%%%%%%%%%%%%%%%%%%%%%%%%%%%%%%%%%%
for $u\in[0,1]$ and $s=1,2,\dots$, where $\ov T_{X,s}(\cdot)\equiv 1-T_{X,s}(\cdot)$. Define, for $s=1,2,\dots,$
\begin{eqnarray}
\mathcal{R}^{-1}_{X,s}(u)&=&1-\mathcal {H}^{-1}_{X,s}(1-u)\label{mre70}
\\&=&\overline{T}_{X,s+1}\left(\overline{T}^{-1}_{X,s}(u)\right).\nonumber
\end{eqnarray}
Note that, for $s=1,2,\dots,$ 
\begin{eqnarray}
\mathcal{R}_{X,s}(u)&=&1-\mathcal {H}_{X,s}(1-u)\label{mre76}
\\&=&\overline{T}_{X,s}\left(\overline{T}^{-1}_{X,s+1}(u)\right).\nonumber
\end{eqnarray}
%%%%%%%%%%%%%%%%%%%%%%%%%%  Lorenz Curve  %%%%%%%%%%%%%%%%%%%%%%%%%%%%%%%%%%%%%%%%%%%%%
The Lorenz curve introduced by Lorenz \cite{lr6}, is basically used to understand the concept of income inequalities in Economics. A brief discussion about Lorenz curve may be found in 
Aaberge \cite{a2}.
The Lorenz curve of $\overline{T}_{X,s}(\cdot)$, denoted by $\mathcal{L}_{X,s}(\cdot)$, is defined as
$$\mathcal{L}_{X,s}(u)=\frac{1}{\wt \mu_{X,s}}\int\limits_0^uT^{-1}_{X,s}(y)dy\;\;\text{for}\;u\in[0,1]\;\text{and}\;s=1,2\dots.$$
In the literature,
the partial orderings with respect to different ageing properties, namely, IFR (Increasing
in Failure Rate), IFRA (Increasing in Failure Rate Average), NBU (New Better than Used),
DMRL (Decreasing in Mean Residual Life), NBUE (New Better than Used in Expectation)
and HNBUE (Harmonically New Better than Used in Expectation) have been defined
and discussed in Bryson and Siddiqui~\cite{bs6}, Barlow and Proschan~\cite{bp1}, Klefsj$\ddot{\rm{o}}$~\cite{kle1} and others.
\\\hspace*{.2in} For the sake of completeness, we reproduce the following definitions of 
generalized ageing classes from Fagiuoli and Pellerey~\cite{fp1}.
%%%%%%%%%%%%%%%%%%%%%%%%%%%%%%%%%%%%%%%%%%%%%%%%%%%%%%%%%%%%%%%%%%%%%%%%%%%%%%%%%%%%%%%%%%%%%%%%%%%%%%%%%%%%%%%%%%%%%%%%%%%%%%%%%%%%%%%%%%%%%%%%%%%%%%%%%%%%%%%%%%%%%%%%%%%%%%%%%%%%%%
\begin{d1} \label{def1-1} For $s=1,2,\ldots$, $X$ is said to be 
\begin{enumerate}
\item[($i$)] s-IFR if $r_{X,s}(x)$ is increasing in $x\ge 0$;
\item[($ii$)] s-IFRA if $\frac{1}{x}\int_0^x r_{X,s}(t)dt$ is increasing in $x > 0$;
\item[($iii$)] s-NBU if $\overline{T}_{X,s}(x+t)\le \overline{T}_{X,s}(x).\overline{T}_{X,s}(t)$ 
for all $x,t\ge 0$;
%\item[($d$)] s-NBUCX if $\int_x^\infty \overline{T}_{X,s}(t+u)du\leqslant\overline{T}_{X,s}(t)
%\int_x^\infty\overline{T}_{X,s}(u)du$, 
%             for all $x,t\geq 0$;
%\item[($e$)] s-NBUCV if $\int_0^x \overline{T}_{X,s}(t+u)du\leqslant\overline{T}_{X,s}(t)
%\int_0^x\overline{T}_{X,s}(u)du$, 
%             for all $x,t\geq 0$;
\item[($iv$)] s-NBUFR if $r_{X,s}(0)\le r_{X,s}(x)$ for all $x\ge 0;$
\item[($v$)] s-NBAFR if $r_{X,s}(0)\le \frac{1}{x}\int_0^x r_{X,s}(x)$ for all $x> 0.$\hfill $\Box$
\end{enumerate}
\end{d1}
%%%%%%%%%%%%%%%%%%%%%%%%%%%%%%%%%%%%%%%%%%%%%%%%%%%%%%%%%%%%%%%%%%%%%%%%%%%%%%%%%%%%%%%%%%%%%%%%%%%%%%%%%%%%%%%%%%%%%%%%%%%%%%%%%%%%%%%%%%%%%%%%%%%%%%%%%%%%%%%%%%%%%%%%%%%%%%%%%%%%%%%%
One can easily verify that each of the following equivalence relations holds:\\\\
\begin{tabular}{lll}
 1-IFR $\Leftrightarrow$ IFR,& 2-IFR $\Leftrightarrow$ DMRL,&
3-IFR $\Leftrightarrow$ DVRL,\\
1-IFRA $\Leftrightarrow$ IFRA,& 2-IFRA $\Leftrightarrow$ DMRLHA,&
1-NBU $\Leftrightarrow$ NBU,\\
 1-NBUFR $\Leftrightarrow$ NBUFR,&
2-NBUFR $\Leftrightarrow$ NBUE, &3-NBUFR $\Leftrightarrow$ NDVRL,\\
%1-NBUCX$\Leftrightarrow$NBUC;&1-NBUCV$\Leftrightarrow$NBU(2);&
1-NBAFR$\Leftrightarrow$NBAFR,&2-NBAFR$\Leftrightarrow$HNBUE.&
\end{tabular}\\
 
For the definitions of DVRL (Decreasing in Variance Residual Life) and NDVRL (Net DVRL) classes
one may refer to Launer~\cite{l6}, 
DMRLHA (Decreasing Mean Residual Life in Harmonic Average)
and NBUFR (New Better than Used in Failure Rate) classes are
discussed in Deshpande et al.~\cite{dks1}, whereas NBAFR (New Better Than Used in Failure Rate Average) is due to Loh~\cite{lw1}.
%whereas HNBUE (Harmonic New Better Than Used in Expectation) class
%is defined in Klefsj$\ddot{\rm{o}}$~\cite{kle1}.\\
%whereas NBUC class is defined in Cao and wang (1991).\\
%\hspace*{.2in} 
% It is not very difficult to see that $X$ is s-IFRA if, and only if,
% \begin{equation}
%  -\frac{\ln \overline{T}_{X,s}(x)}{x}\;{\rm is\; increasing\; in}\; x,\label{eqn1-2}
% \end{equation}
% for $s=1,2,\ldots\cdot$
%%%%%%%%%%%%%%%%%%%%%%%%%%%%%%%%%%%%%%%%%%%%%%%%%%%%%%%%%%%%%%%%%%%%%%%%%%%%%%%%%%%%%%%%%%%%%%%% 
\\\hspace*{.2in} A function $f(\cdot)$ is called star-shaped (resp. antistar-shaped) if $f(x)/x$ is increasing (resp. decreasing) in $x$.
On the other hand, it is called super-additive (resp. sub-additive) if, for all $x,y$, $f(x+y)\geq(resp. \leq~)f(x)+f(y)$.
\\\hspace*{.2in} Let an absolutely continuous nonnegative random variable $Y$ have the respective generalized functions (analogous to the one
defined above for $X$) $\overline{T}_{Y,s}(\cdot),\;\wt \mu_{Y,s}$, $r_{Y,s}(\cdot),$ $\mu_{Y,s}(\cdot),$ $\mathcal{H}_{Y,s}(\cdot)$, $\mathcal{R}_{Y,s}(\cdot)$ and $\mathcal{L}_{Y,s}(\cdot)$.
For the sake of simplicity we write, for $x\geq 0$ and $s=1,2,\dots,$ 
%and $$\mathcal{R}_{Y,s}(x)=\overline{T}_{Y,s+1}(\overline{T}^{-1}_{Y,s}(x)).$$
$$\alpha_s(x)=\overline{T}_{Y,s}^{-1}\left(\overline{T}_{X,s}(x)\right)={T}_{Y,s}^{-1}\left({T}_{X,s}(x)\right).$$
Here, we define and study some more general partial orderings using the generalized ageing properties.
These extend the concepts of the generalized ageing, given in Definition \ref{def1-1}, to
compare the ageing properties of two life distributions. In Sections 2, 3, 4, 5 and 6, we discuss
s-IFR, s-IFRA, s-NBU, s-NBUFR and s-NBAFR orderings, respectively. We give some equivalent representations for each ordering.
We prove that these are all partial orderings. Inter-relations among these orderings are also discussed.
We make a bridge by which one can go from these orderings to generalized ageings, and vice versa.
%We show $-$how these orders are connected with generalized ageings. 
\section{s-IFR Ordering}\label{sec2}
In this section we define s-IFR ordering and study different properties of this ordering.
%%%%%%%%%%%%%%%%%%%%%%%%%%%%%%%%%%%%%%%%%%%%%%%%%%%%  Definition   %%%%%%%%%%%%%%%%%%%%%%%%%%%%%%%%%%%%%%%
\begin{d1} \label{def2-1} For any positive integer $s$, $X$ (or its distribution $F_X$) is said to be more s-IFR than
$Y$ (or its distribution $F_Y$) (written as $F_X\leq_{s-IFR}F_Y$) if
$\alpha_s(x)\;\text{is convex}.$\hfill $\Box$
\end{d1}
%%%%%%%%%%%%%%%%%%%%%%%%%%%%%%%%%  Remark 2.1   %%%%%%%%%%%%%%%%%%%%%%%%%%%%%%%%%%%%%%%%%%%%%%%%%%%%%%%%%%%%%%%%%%%%%%%%%%%%%%%%%%%%%
\begin{r1}\label{rem2-1}
For $s=1$, Definition \ref{def2-1} gives $F_X\leq_{IFR}F_Y$, %and for $s=2$, $F\leq_{DMRL}G$, 
%which has been studied in the literature.
for $s=2$, $F_X\leq_{DMRL}F_Y$, and for $s=3$, we get
$F_X\leq_{DVRL}F_Y$.
%The IFR ordering is known as 
%convex transform ordering (cf. Barlow and Proschan~\cite{bp1}). \hfill $\Box$
\end{r1}
%%%%%%%%%%%%%%%%%%%%%%%%%%%%%%%%%%%%%%%%%%%%%  Lemma %%%%%%%%%%%%%%%%%%%%%%%%%%
\hspace*{0.2 in}The following lemma may be obtained in Marshall and Olkin~(\cite{mo1}, Section $21(f)$, pp. 699-700). 
\begin{l1}\label{l2-1}
 Let $f(\cdot)$ and $g(\cdot)$ be two real-valued continuous functions, and $\zeta(\cdot)$ be a strictly increasing (resp. decreasing) and continuous function defined on the range of $f$ and $g$.
 Then, for any real number $c>0$, $f(x)-cg(x)$ and $\zeta (f(x))-\zeta (cg(x))$ have sign change property in the same (resp. reverse) order, as $x$ traverses from left to right.\hfill $\Box$
\end{l1}

%%%%%%%%%%%%%%%%%%%%%%%%%%%%%%%%%%%%%%%%%%%%%%%%%%%%%%%%%%%%%  Propositon%%%%%%%%%%%%%%%%%%%%%%%%%%%%%%%%%%%%%%%%%%%%%%%%%%%%%%%%%%
\hspace*{0.2 in}In the following two propositions, we give some equivalent representations of the s-IFR ordering. The proof of the first proposition 
can easily be done by using Lemma \ref{l2-1}, or Proposition~2.C.8 of Marshall and Olkin~\cite{mo1}. 
\begin{p1}
 Definition \ref{def2-1} can equivalently be written in one of the following forms:
 \begin{enumerate}
 %\item [$(i)$] $\ov T^{-1}_{Y,s}\ov T_{X,s}(x)$ is convex in
  \item [$(i)$] For any real numbers $a$ and $b$, $\ov T^{-1}_{Y,s}\ov T_{X,s}(x)-(ax+b)$ changes sign at most twice, 
  and if the change of signs occurs twice, they are in the order $+,-,+$, as $x$ traverses from $0$~to~$\infty$.
   \item [$(ii)$] For any real numbers $a$ and $b$, $\ov T_{X,s}(x)-\ov T_{Y,s}(ax+b)$ changes sign at most twice, and if the change of signs occurs twice, they are in the order $-,+,-$,  as $x$ traverses from $0$~to~$\infty$.
    \item [$(iii)$] For any real numbers $a$ and $b$, $\ov T_{X,s}(ax+b)-\ov T_{Y,s}(x)$ changes sign at most twice, and if the change of signs occurs twice, they are in the order $-,+,-$,  as $x$ traverses from $0$~to~$\infty$.
     \item [$(iv)$] For any real numbers $a$ and $b$, $\ov T_{Y,s}(x)-\ov T_{X,s}(ax+b)$ changes sign at most twice, and if the change of signs occurs twice, they are in the order $+,-,+$,  as $x$ traverses from $0$~to~$\infty$.
      \item [$(v)$] For any real numbers $a$ and $b$, $\ov T^{-1}_{X,s}\ov T_{Y,s}(x)-(ax+b)$ changes sign at most twice, and if the change of signs occurs twice, they are in the order $-,+,-$,  as $x$ traverses from $0$~to~$\infty$.
      \item [$(vi)$] $\alpha^{-1}_s(x)$ is concave in $x>0$.\hfill $\Box$
 \end{enumerate}
\end{p1}

%%%%%%%%%%%%%%%%%%%%%%%%%%%%%%%%%%%%%%%%%%%%%%%%%%%%  Proposition  2.1  %%%%%%%%%%%%%%%%%%%%%%%%%%%%%%%%%%%%%%%%%%%%%%%
\begin{p1}
 For $s=2,3,\dots$, Definition \ref{def2-1} can equivalently be written in one of the following forms:
  \begin{enumerate}
  \item [$(i)$] $\frac{r_{X,s}\left(T^{-1}_{X,s}(u)\right)}{r_{Y,s}\left(T^{-1}_{Y,s}(u)\right)}$ is increasing in $u\in[0,1]$.
   \item [$(ii)$] $\frac{\mu_{X,s-1}\left(T^{-1}_{X,s}(u)\right)}{\mu_{Y,s-1}\left(T^{-1}_{Y,s}(u)\right)}$ is decreasing in $u\in[0,1]$.
  \item [$(iii)$] $\frac{\ov T_{Y,s-1}\left(\alpha_s(x)\right)}{\ov T_{Y,s-1}\left(\alpha_{s-1}(x)\right)}$ is decreasing in $x \geq 0$.
  \item [$(iv)$] $\mathcal{R}_{X,s-1}\mathcal{R}^{-1}_{Y,s-1}(u)$ is antistar-shaped in $u\in[0,1]$.
  \item [$(v)$] $\frac{1-\mathcal{H}_{X,s-1}(u)}{1-\mathcal{H}_{Y,s-1}(u)}$ is increasing in $u\in[0,1]$.
  \item [$(vi)$] $\frac{\mathcal{R}_{X,s-1}(u)}{\mathcal{R}_{Y,s-1}(u)}$ is decreasing in $u\in[0,1]$.
  \item [$(vii)$] $\mathcal{R}^{-1}_{Y,s}\mathcal{R}_{X,s}(u)$ is concave in $u\in[0,1]$.
 \end{enumerate}
\end{p1}
{\bf Proof:} 
$F_X\leq_{s-IFR}F_Y$ is equivalent to the fact that
\begin{eqnarray}\label{mre53}
\alpha'_s(x)\;\text{is \;increasing \;in}\;x\geq 0.
\end{eqnarray}
Note that, for $x\geq 0$, 
\begin{eqnarray}
 \alpha'_s(x)&=&\left(\frac{\wt\mu_{Y,s-1}}{\wt\mu_{X,s-1}}\right)\left(\frac{\ov T_{X,s-1}(x)}{\ov T_{Y,s-1}\ov T^{-1}_{Y,s}\left(\ov T_{X,s}(x)\right)}\right)\label{mre51}
 \\&=&\left(\frac{\wt\mu_{Y,s-1}}{\wt \mu_{X,s-1}}\right)\left(\frac{\ov T_{X,s-1}(x)}{\ov T_{X,s}(x)}\right)\left(\frac{\ov T_{X,s}(x)}{\ov T_{Y,s-1}\ov T^{-1}_{Y,s}\left(\ov T_{X,s}(x)\right)}\right)\nonumber
 %\\&=&r_{X,s}(x)\left(\frac{\wt \mu_{Y,s-1}\ov T_{X,s}(x)}{\ov T_{Y,s-1}\ov T^{-1}_{Y,s}\left(\ov T_{X,s}(x)\right)}\right)\nonumber
 \\&=&r_{X,s}(x)\left(\frac{\wt\mu_{Y,s-1}\ov T_{Y,s}\left(\ov T^{-1}_{Y,s}\ov T_{X,s}(x)\right)}{\ov T_{Y,s-1}\left(\ov T^{-1}_{Y,s}\ov T_{X,s}(x)\right)}\right)\nonumber
 %\\&=&\frac{r_{X,s}(x)}{r_{Y,s}\left(\ov T^{-1}_{Y,s}\ov T_{X,s}(x)\right)}\nonumber
 \\&=&\frac{r_{X,s}(x)}{r_{Y,s}\left( T^{-1}_{Y,s}T_{X,s}(x)\right)},\nonumber
\end{eqnarray}
which can equivalently be written as 
\begin{eqnarray}
 \alpha'_s\left( T^{-1}_{X,s}(u)\right)=\frac{r_{X,s}\left(T^{-1}_{X,s}(u)\right)}{r_{Y,s}\left(T^{-1}_{Y,s}(u)\right)}\quad \text{for all}\;u\in[0,1].\label{mre75}
\end{eqnarray}
Thus, the result follows from (\ref{mre53}). This proves $(i)$. Equivalence of $(i)$ and $(ii)$ follows by 
%%%%%%%%%%%%%%%%%% proof b%%%%%%%%%%%%%%%%%%%%%%%%%%%
using (\ref{mre5}) in (\ref{mre75}).
% we get, for all $u\in[0,1]$,
% $$\alpha'_s\left( T^{-1}_{X,s}(u)\right)=\frac{\mu_{Y,s-1}\left(T^{-1}_{Y,s}(u)\right)}{\mu_{X,s-1}\left(T^{-1}_{X,s}(u)\right)}.$$
% Thus, the result follows from ($\ref{mre53}$). 
%%%%%%%%%%%%%%%%%%%%%%%%%%%%%%%%%%%%%%%%%%%%  Proof c %%%%%%%%%%%%%%%%%%%%%%%%%%%%
By noting the fact that   
\begin{eqnarray*}
 \frac{\ov T_{Y,s-1}\left(\alpha_s(x)\right)}{\ov T_{Y,s-1}\left(\alpha_{s-1}(x)\right)}=\left(\frac{\wt\mu_{Y,s-1}}{\wt\mu_{X,s-1}}\right)\frac{1}{\alpha'_s(x)},
\end{eqnarray*}
the equivalence of $(i)$ and $(iii)$ follows from $(\ref{mre53})$. Note that
%%%%%%%%%%%%%%%%%%%%%%%%%%%%%%%%%%%%%%  Proof d %%%%%%%%%%%%%%%%%%%%%%%%
 $$\mathcal{R}_{X,s-1}\mathcal{R}^{-1}_{Y,s-1}(u)\;\text{is antistar-shaped\;in}\;u\in[0,1],$$
 if, and only if, 
$$\frac{\ov T_{X,s-1}\ov T^{-1}_{X,s}\ov T_{Y,s}\ov T^{-1}_{Y,s-1}(u)}{u}\;\text{is \;decreasing\;in}\;u\in[0,1],$$ 
or equivalently, 
$$\frac{\ov T_{X,s-1}(x)}{\ov T_{Y,s-1}\ov T^{-1}_{Y,s}\left(\ov T_{X,s}(x)\right)}\;\text{is \;increasing\;in}\; x\geq 0.$$
Thus, the equivalence of $(i)$ and $(iv)$ follows from (\ref{mre53}) and (\ref{mre51}).
%%%%%%%%%%%%%%%%%%%%%%%%%%%%%%%%%%%%%%%%%%%%%%%%%%%%  proof e%%%%%%%%%%%%%%%%%%%
The equivalence of $(i)$ and $(v)$ follows from $(\ref{mre51})$ and using the fact that
\begin{eqnarray*}
 \frac{1-\mathcal{H}_{X,s-1}(u)}{1-\mathcal{H}_{Y,s-1}(u)}&=&\frac{\ov T_{X,s-1}\left(T^{-1}_{X,s}(u)\right)}{\ov T_{Y,s-1} \left(T^{-1}_{Y,s}(u)\right)}.
\end{eqnarray*}
Equivalence of $(v)$ and $(vi)$ follows from (\ref{mre76}).
 We write $\Upsilon_{X,s}(u)=\mathcal{R}'_{X,s}(u)$ and $\Upsilon_{Y,s}(u)=\mathcal{R}'_{Y,s}(u)$ for $u\in[0,1]$. Then, we have, for all $u\in[0,1]$,
$$\Upsilon_{X,s}(u)=\left(\frac{\wt \mu_{X,s}}{\wt \mu_{X,s-1}}\right)\left(\frac{\ov T_{X,s-1}\left(\ov T^{-1}_{X,s+1}(u)\right)}{\ov T_{X,s}\left(\ov T^{-1}_{X,s+1}(u)\right)}\right),$$
which gives
\begin{eqnarray*}
 \Upsilon_{X,s}\left(\mathcal{R}^{-1}_{X,s}(u)\right)=\left(\frac{\wt\mu_{X,s}}{\wt\mu_{X,s-1}}\right)\left(\frac{\ov T_{X,s-1}\left(\ov T^{-1}_{X,s}(u)\right)}{u}\right).
\end{eqnarray*}
So, on using (\ref{mre51}) we have, for all $u\in[0,1]$,
\begin{eqnarray}\label{cr6}
 \frac{\Upsilon_{X,s}\left(\mathcal{R}^{-1}_{X,s}(u)\right)}{\Upsilon_{Y,s}\left(\mathcal{R}^{-1}_{Y,s}(u)\right)}
=\left(\frac{\wt\mu_{X,s}}{\wt\mu_{Y,s}}\right)\alpha'_{s}\left(\ov T^{-1}_{X,s}(u)\right).
 \end{eqnarray}
 Thus, (\ref{mre53}) can equivalently be written as 
 $$\frac{\Upsilon_{X,s}\left(\mathcal{R}^{-1}_{X,s}(u)\right)}{\Upsilon_{Y,s}\left(\mathcal{R}^{-1}_{Y,s}(u)\right)}\;\text{is decreasing in}\;u\in[0,1],$$
 or equivalently,
 $$\frac{\Upsilon_{X,s}(u)}{\Upsilon_{Y,s}\left(\mathcal{R}^{-1}_{Y,s}\mathcal{R}_{X,s}(u)\right)}\;\text{is decreasing in}\;u\in[0,1].$$
 This means that
 $$\frac{d}{du}\left(\mathcal{R}^{-1}_{Y,s}\mathcal{R}_{X,s}(u)\right)\;\text{is decreasing in}\;u\in [0,1],$$
 or equivalently,
 $$\mathcal{R}^{-1}_{Y,s}\mathcal{R}_{X,s}(u)\;\text{is concave in}\;u\in [0,1].$$
 This gives the equivalence of $(i)$ and $(vii)$.
\hfill$\Box$
%%%%%%%%%%%%%%%%%%%%%%%%%%%%%%%%%%%%%   begin Corollary   %%%%%%%%%%%%%%%%%%%%%%%%%%%%
\begin{r1}
 For $s=1$, Definition \ref{def2-1} can equivalently be written in one of the following forms:
  \begin{enumerate}
  \item [$(i)$] $\frac{r_{X,1}\left(F^{-1}_{X}(u)\right)}{r_{Y,1}\left(F^{-1}_{Y}(u)\right)}$ is increasing in $u\in[0,1]$.
  \item [$(ii)$] $\mathcal{R}^{-1}_{Y,1}\mathcal{R}_{X,1}(u)$ is concave in $u\in[0,1]$.\hfill$\Box$
  \end{enumerate}
\end{r1}
%%%%%%%%%%%%%%%%%%%%%%%%%%%%%%%%%%%%%%%%%  Definition 2.2   %%%%%%%%%%%%%%%%%%%%%%%%%%%%%%%%%%%%%%%%%%%%%%%%%%%%%%%%%%%%%%%%%%%%%%%%%%%%%%%%%%%%%%%
\begin{d1}\label{def2-2} Two distribution functions $F_X,F_Y(\in{\cal F})$ are said to
be equivalent $(F_X\sim F_Y)$ if there exists a $\theta>0$ such that $F_X(x)=F_Y(\theta x)$ for all $x\ge 0.$\hfill$\Box$
\end{d1}
%%%%%%%%%%%%%%%%%%%%%%%%%%%%%%%%%%%%%%%%%%%%%%%%%%%%%%%%%%%%%%%%%%%%%%%%%%%%%%%%%%%%%%%%%%%%%%%%%%%%%%%%%%%%%%%%%%%%%%%%%%%%%%%%%%%%%%%%%%%%%%%%%%%%%%%%%%%%%%%%%%%%%%%%%%%%%%%%%%%%%%%%%%%
\hspace*{.2in} Following are a few lemmas to be used in proving the upcoming theorems.
\begin{l1}\label{lem2-1} If $F_X\sim F_Y$, then $\overline{T}_{X,s}(x)=\overline{T}_{Y,s}(\theta x)$
for some $\theta >0$ and all $x\ge 0$, and $s=1,2,\ldots\cdot$
\end{l1}
{\bf Proof:} $F_X\sim F_Y$ if, and only if $\overline{F}_X(x)=\overline{F}_Y(\theta x)$ for 
some $\theta >0$ and for all $x\ge 0$. Thus the result is true for $s=1$. Suppose the result holds for $s$. Then
$$\overline{T}_{X,s+1}(x)=\frac{1}{\wt \mu_{X,s}}\int_x^\infty \overline{T}_{X,s}(u)du.$$
Further
\begin{eqnarray*}
\wt \mu_{X,s}&=&\int_0^\infty \overline{T}_{X,s}(u)du\\
&=&\int_0^\infty \overline{T}_{Y,s}(\theta u)du\\
&=&\frac{\wt\mu_{Y,s}}{\theta}.
\end{eqnarray*}
The second equality follows from the hypothesis. Hence
\begin{eqnarray*}
\overline{T}_{X,s+1}(x)&=&\frac{\theta}{\wt\mu_{Y,s}}\int_x^\infty \overline{T}_{Y,s}(\theta u)du\\
%&=& \frac{1}{\wt \mu_{Y,s}}\int_{\theta x}^\infty \overline{T}_{Y,s}(u)du\\
&=& \overline{T}_{Y,s+1}(\theta x).
\end{eqnarray*}
Hence, by induction, the result is established.\hfill $\Box$
\\\hspace*{0.2 in} Following lemma follows from the definition of $r_{X,s}(\cdot)$ and Lemma~\ref{lem2-1}.
%%%%%%%%%%%%%%%%%%%%%%%%%%%%%%%%%%%%%%%%%%%%%%%%%%%%%%%%%%%%%%%%%%%%%%%%%%%%%%%%%%%%%%%%%%%%%%%%%%%%%%%%%%%%%%%%%%%%%%%%%%%%%%%%%%%%%%%%%%%%%%%%%%%%%%%%%%%%%%%%%%%%%%%%%%%%%%%%%%%%%
\begin{l1}\label{lem2-2} If $F_X\sim F_Y$, then there exists a $\theta>0$ such that, for all $x\ge 0$ and
$s=1,2,\ldots,$
$$r_{X,s}(x)=\theta r_{Y,s}(\theta x).$$
\end{l1}\hfill $\Box$
\\\hspace*{.2in} The following lemma gives the converse of Lemma \ref{lem2-1}.
%%%%%%%%%%%%%%%%%%%%%%%%%%%%%%%%%%%%%%%%%%%%%%%%%%%%%%%%%%%%%%%%%%%%%%%%%%%%%%%%%%%%%%%%%%%%%%%%%%%%%%%%%%%%%%%%%%%%%%%%%%%%%%%%%%%%%%%%%%%%%%%%%%%%%%%%%%%%%%%%%%%%%%%%%%%%%%%%%%%%%%%%%
\begin{l1}\label{lem2-4} If $\overline{T}_{X,s}(x)=\overline{T}_{Y,s}(\theta x)$ for some
$\theta>0$, some $s=1,2,\ldots$, and all $x\ge 0$, then $\overline{F}_X(x)=\overline{F}_Y(\theta x)$ for
all $x$.
\end{l1}
{\bf Proof:} Let us fix $s\geq 2$ because for $s=1$, it is trivial. 
Then $\overline{T}_{X,s}(x)=\overline{T}_{Y,s}(\theta x)$ for all $x\ge 0$ gives,
by ($\ref{eqn1-1}$),
$$\frac{\int_x^\infty\overline{T}_{X,s-1}(u)du}{\wt\mu_{X,s-1}}=
\frac{\int_{\theta x}^\infty\overline{T}_{Y,s-1}(u)du}{\wt \mu_{Y,s-1}}\quad\text{for all}\;x\geq 0.$$
Taking derivative with respect to $x$ on both sides of the above expression, we get, for all $x\ge 0$,
\begin{equation}\label{eqn2-3}
\frac{\overline{T}_{X,s-1}(x)}{\wt \mu_{X,s-1}}=\theta\frac{\overline{T}_{Y,s-1}
(\theta x)}{\wt \mu_{Y,s-1}}\quad{\rm for \;
all}\; 
x\ge 0.
\end{equation}
Putting $x=0$ in ($\ref{eqn2-3}$), we get $\wt \mu_{Y,s-1}/\wt \mu_{X,s-1}=\theta$. Hence
($\ref{eqn2-3}$) becomes
$$\overline{T}_{X,s-1}(x)= \overline{T}_{Y,s-1}(\theta x)\quad\text{for all}\; x\ge0.$$
Proceeding in this line, we get $\overline{F}_X(x)=\overline{F}_Y(\theta x)$ for all
$x\ge 0.$ \hfill$ \Box$
%%%%%%%%%%%%%%%%%%%%%%%%%%%%%%%%%%%%%%%%%%%%%%%%%%%%%%%%%%%%%%%%%%%%%%%%%%%%%%%%%%%%%%%%%%%%%%%%%%%%%%%%%%%%%%%%%%%%%%%%%%%%%%%%%%%%%%%%%%%%%%%%%%%%%%%%%%%%%%%%%%%%%%%%%%%%%%%%%%%%%%%%5555
\\\hspace*{0.2 in} The following two lemmas are easy to prove.
\begin{l1}\label{lem2-5}
 Let $f(\cdot)$ and $g(\cdot)$ be two nonnegative, increasing, and convex functions. Then $f\left(g(\cdot)\right)$ is convex.\hfill $\Box$
\end{l1}
%%%%%%%%%%%%%%%%%%%%%%%%%%%%%%%%%%%%%%%%%%%%%%%%%%%%%%%%%%%%%%%%%%%%%%%%%%%%%%%%%%%%%%%%%%%%%%%%%%%%%%%%%%%%%%%%%%%%%%%%%%%%%%%%%%%%%%%%%%%%%%%%%%%%%%%%%%%%%%%%%%%%%%%%%%%%%%%%%%%%%%%%%%
\begin{l1}\label{lem2-6}
 Let $f(\cdot)$ be a nonnegative, increasing and convex function. Then $f^{-1}(\cdot)$ is concave.\hfill $\Box$
\end{l1}
%%%%%%%%%%%%%%%%%%%%%%%%%%%%%%%%%%%%%%%%%%%%%%%%%%%%%%%%%%%%%%%%%%%%%%%%%%%%%%%%%%%%%%%%%%%%%%%%%%%%%%%%%%%%%%%%%%%%%%%%%%%%%%%%%%%%%%%%%%%%%%%%%%%%%%%%%%%%%%%%%%%%%%%%%%%%%%%%%%%%%%%%%%
% \begin{r1}\label{remaa}
% In the proof below, we have used the fact that 
% $\overline{T}_{Y,s}^{-1}\overline{T}_{X,s}(\cdot)\equiv T_{Y,s}^{-1}T_{X,s}(\cdot)$.
% \end{r1}
%%%%%%%%%%%%%%%%%%%%%%%%%%%%%%%%%%%%%%%%%%%%%%%%%%%%%%%%%%%%%%%%%%%%%%%%%%%%%%%%%%%%%%%%%%%%%%%%%%%%%%%%%%%%%%%%%%%%%%%%%%%%%%%%%%%%%%%%%%%%%%%%%%%%%%%%%%%%%%%%%%%%%%%%%%%%%%%%%%%%%%%%%
\hspace*{0.2 in}The following theorem shows that s-IFR ordering is a partial ordering.
\begin{t1}\label{th2-1} The relationship $F_X\leq_{s-IFR}F_Y$ is a partial
ordering of the equivalence classes of $\cal{F}$.
\end{t1}
{\bf Proof:} ($i$) That s-IFR ordering is reflexive, is trivial.\\
($ii$)  $F_X\leq_{s-IFR}F_Y$ gives that $T_{Y,s}^{-1}\left(T_{X,s}(x)\right)$ is convex, which, by Lemma \ref{lem2-6}, 
reduces to the fact that 
\begin{equation}\label{aa}
 T_{X,s}^{-1}\left(T_{Y,s}(x)\right)\;{\rm is\; concave}.
\end{equation}
 Further, $F_Y\leq_{s-IFR}F_X$ gives that
\begin{equation}\label{ab}
 T_{X,s}^{-1}\left(T_{Y,s}(x)\right)\;{\rm is\; convex}.
\end{equation}
Combining (\ref{aa}) and (\ref{ab}), we get
$$T_{X,s}^{-1}\left(T_{Y,s}(x)\right)=\alpha+\beta x,$$
for some constants $\alpha$ and $\beta$. Now, by evaluating the above expression at $x=0$, we get $\alpha=0$. Hence, we have 
$$T_{X,s}^{-1}\left(T_{Y,s}(x)\right)=\beta x,$$
which, by Lemma \ref{lem2-4}, gives $F_X\sim F_Y$.\\
($iii$) On using Lemma \ref{lem2-5}, one can easily see that $s$-IFR ordering is transitive.\hfill $\Box$
%%%%%%%%%%%%%%%%%%%%%%%%%%%%%%%%%%%%%%%%%%%%%%%%%%%%%%%%%%%%%%%%%%%%%%%%%%%%%%%%%%%%%%%%%%%%%%%%%%%%%%%%%%%%%%%    Theorem   %%%%%%%%%%%%%%%%%%%%%%%%%%%%%%%%%%%%%%%%%%%%%%%%%%%
% \begin{t1}
%  \begin{enumerate}
%   \item [$(i)$] $X\leq_{s-IFR}Y$ if and only if $aX\leq_{s-IFR}aY$ for all $a>0.$
%   \item [$(ii)$] $X\leq_{s-IFR}Y$ if and only if $X^r\leq_{s-IFR}Y^r$ for all $r>0$.
%  \end{enumerate}
% \end{t1}
%%%%%%%%%%%%%%%%%%%%%%%%%%%%%%%%%%%%%%%%%%%%%%%%%%%%%%%%%%%%%%%%%%%%%%%%%%%%%%%%%%%%%%%%%%%%%%%%%%%%%%%%%%%%%%%%%%%%%%%%%%%%%%%%%%%%%%%%%%%%%%%%%%%%%%%%%%%%%%%%%%%%%%%%%%%%%%%%%%%%%%%%%
\\\hspace*{.2in} The following lemma can be easily verified.
\begin{l1}\label{lem2-3} Let $X\sim \overline{F}_X(x)=e^{-\lambda x}$. Then, for $s=1,2,\ldots$,
\begin{enumerate}
\item[($i$)] $r_{X,s}(x)=\lambda $;
\item[($ii$)] $\overline{T}_{X,s}(x)=e^{-\lambda x}.$\hfill $\Box$
\end{enumerate}
\end{l1}
%%%%%%%%%%%%%%%%%%%%%%%%%%%%%%%%%%%%%%%%%%%%%%%%%%%%%%%%%%%%%%%%%%%%%%%%%%%%%%%%%%%%%%%%%%%%%%%%%%%%%%%%%%%%%%%%%%%%%%%%%%%%%%%%%%%%%%%%%%%%%%%%%%%%%%%%%%%%%%%%%%%%%%%%%%%%%%%%%%%%%%%%
\hspace*{0.2 in}The following theorem shows that a random variable $X$ is smaller than exponential distribution in s-IFR ordering if, and only if,
$X$ has s-IFR distribution. 
\begin{t1}\label{th2-2}
If $\overline{F}_Y(x)=\exp (-\lambda x)$, then $F_X\leq_{s-IFR}F_Y$ if, and only if, $F_X$
is s-IFR.
\end{t1}
{\bf Proof:} By Lemma \ref{lem2-3}, $F_X\leq_{s-IFR}F_Y$ is equivalent to saying that
$$\ln \left(\overline{T}_{X,s}(x)\right)\;{\rm is\;concave},$$
or equivalently,
$$r_{X,s}(x)\;{\rm is\; increasing\;in}\;x\ge 0,$$
giving that  $X$ is s-IFR.\hfill $\Box$
%%%%%%%%%%%%%%%%%%%%%%%%%%%%%%%%%%%%%%%%%%%%%%%%%%%%%%%%%%%%%%%%%%%%%%%%%%%%%%%%%%%%%%%%%%%%%%%%%%%%%%%%%%%%%%%%%%%%%%%%%%%%%%%%%%%%%%%%%%%%%%%%%%%%%%%%%%%%%%%%%%%%%%%%%%%%%
% \begin{t1}\label{th2-3}
% If $F_X\leq_{s-IFR}F_Y$, then $F_X\leq_{(s+1)-IFR}F_Y$.
% \end{t1}
% {\bf Proof:} {\bf To be Constructed.}
\section{s-IFRA Ordering}\label{sec3}
We start this section with the following definition.
\begin{d1} \label{def3-1}For any positive integer $s$,
$X$ (or its distribution $F_X$) is said to be more s-IFRA than Y (or its distribution $F_Y$) (written as $F_X\leq_{s-IFRA}F_Y$) if
$\alpha_s(x)\;\text{is star-shaped}$.
\end{d1}\hfill $\Box$
%%%%%%%%%%%%%%%%%%%%%%%%%%%%%%%%%%%%%%%%%%%%%%%%%%%%%%%%%%%%%%%%%%%%%%%%%%%%%%%%%%%%%%%%%%%%%%%%%%%%%%%%%%%%%%%%%%%%%%%%%%%%%%%%%%%%%%%%%%%%%%%%%%%%%%%%%%%%%%%%%%%%%%%%%%%%%%%%%%%
\begin{r1} \label{rem3-1}
For $s=1$, the above definition gives $F_X\leq_{IFRA}F_Y$.
%or equivalently,
%$F_Y^{-1}F_X(x)$ is star-shaped. This is discussed in Barlow and Proschan~\cite{bp1}.\hfill $\Box$
\end{r1}
%%%%%%%%%%%%%%%%%%%%%%%%%%%%%%%%%%%%%%%%%%%%%%%%%%%%%%%%%%%%%  Propositon%%%%%%%%%%%%%%%%%%%%%%%%%%%%%%%%%%%%%%%%%%%%%%%%%%%%%%%%%%
\hspace*{0.2 in}Below we give some equivalent representations of s-IFRA ordering. The first proposition can easily be proved by using Lemma \ref{l2-1}. 
\begin{p1}\label{p3-1}
 Definition \ref{def3-1} can equivalently be written in one of the following forms:
 \begin{enumerate}
  \item [$(i)$] For any real number $a$, $\ov T^{-1}_{Y,s}\ov T_{X,s}(x)-ax$ changes sign at most once, and if the change of sign does occur, it is in the order $-,+$, as $x$ traverses from $0$ to $\infty$.
   \item [$(ii)$] For any real number $a$, $\ov T_{X,s}(x)-\ov T_{Y,s}(ax)$ changes sign at most once, and if the change of sign does occur, it is in the order $+,-$,  as $x$ traverses from $0$ to $\infty$.
    \item [$(iii)$] For any real number $a$, $\ov T_{X,s}(ax)-\ov T_{Y,s}(x)$ changes sign at most once, and if the change of sign does occur, it is in the order $+,-$,  as $x$ traverses from $0$ to $\infty$.
     \item [$(iv)$] For any real number $a$, $\ov T_{Y,s}(x)-\ov T_{X,s}(ax)$ changes sign at most once, and if the change of sign does occur, it is in the order $-,+$,  as $x$ traverses from $0$ to $\infty$.
      \item [$(v)$] For any real number $a$, $\ov T^{-1}_{X,s}\ov T_{Y,s}(x)-ax$ changes sign at most once, and if the change of sign does occur, it is in the order $+,-$,  as $x$ traverses from $0$ to $\infty$.
      \item [$(vi)$] $\alpha^{-1}_s(x)$ is antistar-shaped in $x>0$.\hfill $\Box$
 \end{enumerate}
\end{p1}

%%%%%%%%%%%%%%%%%%%%%%%%%%%%%%%%%%%%%%%%%%%  Proposition 3.1   %%%%%%%%%%%%%%%%%%%%%%%%%%%%%%%%%%%%%%%%%%%%%%%%%%%%%%%
\begin{p1}\label{pe3-1}
 For $s=2,3,\dots$, Definition \ref{def3-1} can equivalently be written in one of the following forms:
  \begin{enumerate}
  \item [$(i)$] $\frac{\ov T^{-1}_{Y,s}(u)}{\ov T^{-1}_{X,s}(u)}\;\text{is decreasing in}\;u\in[0,1].$ 
  \item [$(ii)$] $\frac{r_{X,s}\left(T^{-1}_{X,s}(u)\right)}{r_{Y,s}\left(T^{-1}_{Y,s}(u)\right)}\geq \frac{T^{-1}_{Y,s}(u)}{T^{-1}_{X,s}(u)}\;\text{for all} \;u\in [0,1].$
   \item [$(iii)$] $\frac{\mu_{Y,s-1}\left(T^{-1}_{Y,s}(u)\right)}{\mu_{X,s-1}\left(T^{-1}_{X,s}(u)\right)}\geq \frac{T^{-1}_{Y,s}(u)}{T^{-1}_{X,s}(u)}$ for all $u\in [0,1]$.
  \item [$(iv)$] $\frac{\ov T_{Y,s-1}\left(\alpha_{s-1}(x)\right)}{\ov T_{Y,s-1}\left(\alpha_{s}(x)\right)}\geq \left(\frac{\wt \mu_{X,s-1}}{\wt \mu_{Y,s-1}}\right)\left(\frac{\alpha_s(x)}{x}\right)$ for all $x \geq 0$.
  \item [$(v)$] $\frac{\mathcal{R}_{X,s-1}(u)}{\mathcal{R}_{Y,s-1}(u)}\geq \left(\frac{\wt \mu_{X,s-1}}{\wt \mu_{Y,s-1}}\right)\left(\frac{\ov T^{-1}_{Y,s}(u)}{\ov T^{-1}_{X,s}(u)}\right)$ for all $u\in [0,1]$.
 \item [$(vi)$] $\frac{1-\mathcal{H}_{X,s-1}(u)}{1-\mathcal{H}_{Y,s-1}(u)}\geq \left(\frac{\wt \mu_{X,s-1}}{\wt \mu_{Y,s-1}}\right)\left(\frac{ T^{-1}_{Y,s}(u)}{ T^{-1}_{X,s}(u)}\right)$ for all $u\in [0,1]$.
 \end{enumerate}
\end{p1}
{\bf Proof:} The proof of $(i)$ follows from definition. Again, $(i)$ can equivalently be written as
\begin{eqnarray}\label{mre54}
 \left(\frac{\wt \mu_{Y,s-1}}{\wt \mu_{X,s-1}}\right)\left(\frac{\ov T_{X,s-1}(x)}{\ov T_{Y,s-1}\ov T^{-1}_{Y,s}\left(\ov T_{X,s}(x)\right)}\right)\geq \frac{\ov T^{-1}_{Y,s}\ov T_{X,s}(x)}{x}.
\end{eqnarray}
The above inequality holds if, and only if, for all $x\geq 0$,
$$\frac{r_{X,s}\left(T^{-1}_{X,s}(u)\right)}{r_{Y,s}\left(T^{-1}_{Y,s}(u)\right)}\geq \frac{T^{-1}_{Y,s}(u)}{T^{-1}_{X,s}(u)}\;\text{for all} \;u\in [0,1],$$
which is $(ii)$. Equivalence of $(ii)$ and $(iii)$ follows from (\ref{mre5}).
%%%%%%%%%%%%%%%%%%%%%%%%%%%%%%%%%%%%%  Proof b%%%%%%%%%%
 Note that 
\begin{eqnarray*}
 \frac{\ov T_{Y,s-1}\left(\alpha_{s-1}(x)\right)}{\ov T_{Y,s-1}\left(\alpha_{s}(x)\right)}&=&\frac{\ov T_{X,s-1}(x)}{\ov T_{Y,s-1}\ov T^{-1}_{Y,s}\left(\ov T_{X,s}(x)\right)}
 \\&\geq& \left(\frac{\wt \mu_{X,s-1}}{\wt \mu_{Y,s-1}}\right)\left(\frac{\alpha_s(x)}{x}\right),
\end{eqnarray*}
where the inequality follows from (\ref{mre54}). This gives the equivalence of $(i)$ and $(iv)$. 
On using (\ref{mre54}), the equivalence of $(iv)$ and $(v)$ follows.
 For $u\in [0,1],$  
\begin{eqnarray*}
 \frac{1-\mathcal{H}_{X,s-1}(u)}{1-\mathcal{H}_{Y,s-1}(u)}&=&\frac{\ov T_{X,s-1}\left(T^{-1}_{X,s}(u)\right)}{\ov T_{Y,s-1}\left( T^{-1}_{Y,s}(u)\right)}
 \\&\geq& \left(\frac{\wt \mu_{X,s-1}}{\wt \mu_{Y,s-1}}\right)\left(\frac{T^{-1}_{Y,s}(u)}{T^{-1}_{X,s}(u)}\right),
\end{eqnarray*}
where the inequality follows from (\ref{mre54}). Hence, the equivalence of $(i)$ and $(vi)$ follows.\hfill$\Box$
\begin{r1}
 For $s=1$, Definition \ref{def3-1} can equivalently be written in one of the following forms:
 \begin{enumerate}
  \item [$(i)$] $\frac{\bar F^{-1}_{Y}(u)}{\bar F^{-1}_{X}(u)}\;\text{is decreasing in}\;u\in[0,1].$ 
  \item [$(ii)$] $\frac{r_{X,1}\left(F^{-1}_{X}(u)\right)}{r_{Y,1}\left(F^{-1}_{Y}(u)\right)}\geq \frac{F^{-1}_{Y}(u)}{F^{-1}_{X}(u)}\;\text{for all} \;u\in [0,1].$\hfill$\Box$
\end{enumerate}
  \end{r1}
%%%%%%%%%%%%%%%%%%%%%%%%%%%%%%%%%%%%%%%%%%%%%%%%%%%%%%%  Theorem   %%%%%%%%%%%%%%%%%%%%%%%%%%%%%%%%%%%%%%%%%%%%%%%%%%%%%%%%%%%%%%%
\hspace*{0.2 in}The following theorem gives some equivalent characterization of s-IFRA ordering. 
\begin{t1}
 The following statements are equivalent:
 \begin{enumerate}
  \item [$(i)$] $F_{X}\leq_{s-IFRA}F_Y$.
  \item [$(ii)$] For all functions $\alpha(\cdot)$ and $\beta(\cdot)$, such that $\alpha(\cdot)$ is nonnegative and $\alpha(\cdot)$ and $\alpha(\cdot)/\beta(\cdot)$ are decreasing, 
  and such that $\int\limits_0^1\alpha(u)d T^{-1}_{X,s}(u)<\infty$, and $\int\limits_0^1\alpha(u)dT^{-1}_{Y,s}(u)<\infty$, \\$0\neq \int\limits_0^1\beta(u)dT^{-1}_{X,s}(u)<\infty$,
  and $0\neq\int\limits_0^1\beta(u)dT^{-1}_{Y,s}(u)<\infty$, we have
  $$\frac{\int\limits_0^1\alpha(u)dT^{-1}_{Y,s}(u)}{\int\limits_0^1\alpha(u)dT^{-1}_{Y,s}(u)}\leq \frac{\int\limits_0^1\beta(u)dT^{-1}_{X,s}(u)}{\int\limits_0^1\beta(u)dT^{-1}_{X,s}(u)}.$$
  \item [$(iii)$] For any increasing functions $a(\cdot)$ and $b(\cdot)$ such that $b(\cdot)$ is nonnegative, if \\$\int\limits_0^1a(u)b(u)dT^{-1}_{X,s}(u)=0$, 
  then $\int\limits_0^1a(u)b(u)dT^{-1}_{Y,s}(u)\leq 0$.
 \end{enumerate}
\end{t1}
{\bf Proof:} The proof follows from Theorem 4.B.10 of Shaked and Shanthikumar \cite{shak1} by noting the fact that $T_{X,s}$
and $T_{Y,s}$ are playing the role of $F$ and $G$, respectively.\hfill$\Box$
%%%%%%%%%%%%%%%%%%%%%%%%%%%%%%%%%%%%%%%%%%%%%%%%%%%%%%%%%%%%%%%%%%%%%%%%%%%%%%%%%%%%  Lemma 3.2  %%%%%%%%%%%%%%%%%%%%%%%%%%%%%%%%%%%%%%%%%%%%%%%%%%%%%%%%%%%%%%%%%%%%%%%%%%%%%%%%%%
\\\hspace*{0.2 in}Below we give two lemmas to be used in the upcoming theorem. The proofs are omitted.
\begin{l1}\label{lem3-1}
 Let $f(\cdot)$ and $g(\cdot)$ be two nonnegative, increasing, and star-shaped functions. Then $f\left(g(\cdot)\right)$ is star-shaped.\hfill$\Box$
\end{l1}
% {\bf Proof:} Given that $f(x)/x$ is increasing, and $g(x)/x$ is increasing. Now the proof follows by writing 
% $f(g(x))/x=[f(g(x))]/g(x)[g(x)/x]$.\hfill$\Box$
%%%%%%%%%%%%%%%%%%%%%%%%%%%%%%%%%%%%%%%%%%%%%%%%%%%%%%%%%%%%%%%%%%%%%%%%%%%%%%%%%%%%%%%%%%%%%%%%%%%%%%%%%%%%%%%%%%%%%%%%%%%%%%%%%%%%%%%%%%%%%%%%%%%%%%%%%%%%%%%%%%%%%%%%%%%%%%%%%%%%%%%%%%
\begin{l1}\label{lem3-2}
 Let $f(\cdot)$ be a nonnegative, increasing, and star-shaped function. Then $f^{-1}(\cdot)$ is antistar-shaped.\hfill$\Box$
\end{l1}
% {\bf Proof:} Given that $f(x)/x$ is increasing. This means
% \begin{equation}\label{ac}
%  xf'(x)\geq f(x).
% \end{equation}
% Now 
% \begin{eqnarray*}
%  \frac{d}{dx}\left(\frac{f^{-1}(x)}{x}\right)&=&\frac{x}{f'(f^{-1}(x))}-f^{-1}(x)\\
%  &=&\frac{f(f^{-1}(x))}{f'(f^{-1}(x))}-f^{-1}(x)\\
%  &\leq &f^{-1}(x)-f^{-1}(x)\\
%  &=&0.
% \end{eqnarray*}
% The above inequality follows from (\ref{ac}).\hfill$\Box$
%%%%%%%%%%%%%%%%%%%%%%%%%%%%%%%%%%%%%%%%%%%%%%%%%%%%%%%%%%%%%%%%%%%%%%%%%%%%%%%%%%%%%%%%%%%%%%%%%%%%%%%%%%%%%%%%%%%%%%%%%%%%%%%%%%%%%%%%%%%%%%%%%%%%%%%%%%%%%%%%%%%%%%%%%%%%%%%%%%%%%
\hspace*{0.2 in}Below we show that s-IFRA ordering is a partial ordering.
\begin{t1}\label{th3-1} The relationship $F_X\leq_{s-IFRA}F_Y$ is a partial ordering of
the equivalence classes of $\cal{F}$.
\end{t1}
{\bf Proof:} ($i$) It is trivial to show that $s$-IFRA ordering is reflexive.\\
($ii$) $F_X\leq_{s-IFRA}F_Y$ gives that $T_{Y,s}^{-1}(T_{X,s}(x))$ is star-shaped, which, by Lemma \ref{lem3-2},
reduces to the fact that 
\begin{eqnarray}
T_{X,s}^{-1}(T_{Y,s}(x))\;\text{is antistar-shaped}.\label{fr11}
\end{eqnarray}
Further, $F_Y\leq_{s-IFRA}F_X$ gives that
\begin{eqnarray}
T_{X,s}^{-1}(T_{Y,s}(x))\;\text{is star-shaped}.\label{fr22}
\end{eqnarray}
Combining (\ref{fr11}) and (\ref{fr22}), we have
$$T_{X,s}^{-1}(T_{Y,s}(x))=\theta x,$$
for some constant $\theta$. This, by Lemma \ref{lem2-4}, gives $F_X\sim F_Y$.\\
($iii$) By Lemms \ref{lem3-1}, we have that the s-IFRA ordering is transitive.\hfill$\Box$
%%%%%%%%%%%%%%%%%%%%%%%%%%%%%%%%%%%%%%%%%%%%%%%%%%%%%%%%%%%%%%%%%%%%%%%%%%%%%%%%%%%%%%%%%%%%%%%%%%%%%%%%%%%%%%%%%%%%%%%%%%%%%%%%%%%%%%%%%%%%%%%%%%%%%%%%%%%%%%%%%%%%%%%%%%%%%%%%%%
\\\hspace*{0.2 in}The following theorem is a bridge between s-IFRA ordering and s-IFRA ageing.
\begin{t1}\label{th3-2} If $\overline{F}_Y(x)=e^{-\lambda x},\;\lambda>0$, then
$$F_X\leq_{IFRA}F_Y\;\text{if, and only if,}\;F_X\;\text{is s-IFRA}.$$
\end{t1}
{\bf Proof:} The proof follows from Definition $\ref{def3-1}$ and Lemma \ref{lem2-3}.\hfill$\Box$
%%%%%%%%%%%%%%%%%%%%%%%%%%%%%%%%%%%%%%%%%%%%%%%%%%%%%%%%%%%%%%%%%%%%%%%%%%%%%%%%%%%%%%%%%%%%%%%%%%%%%%%%%%%%%%%%%%%%%%%%%%%%%%%%%%%%%%%%%%%%%%%%%%%%%%%%%%%%%%%%%%%%%%%%%%%%%%%%%%%%%%%%%55
\\\hspace*{.2in} Since every convex function is star-shaped, we have the following theorem.
\begin{t1} If
 $F_X\leq_{s-IFR}F_Y$, then $F_X\leq_{s-IFRA}F_Y.$\hfill$\Box$
\end{t1}
%%%%%%%%%%%%%%%%%%%%%%%%%%%%%%%%%%%%%%%%%%%%%%%%%%%%%%%%%%%%%%  Theorem   %%%%%%%%%%%%%%%%%%%%%%%%%%%%%%%%%%%%%%%%%%%%
% \begin{t1}
%  Let $X$ be a nonnegative random variable that is not degenerate at $0$, and let $g$ and $h$ be nonnegative increasing functions, defined on $[0,\infty)$. Assume that $g(x)>0$ and $h(x)>0$
%  for all $x>0$. If $h(x)/g(x)$ is increasing in $x>0$, then $g(X)\leq_{s-IFRA} h(X)$.
% \end{t1}
% {\bf Proof:} $g(X)\leq_{s-IFRA} h(X)$ holds if, and only if 
% $$\frac{\ov T^{-1}_{h(X),s}(x)}{\ov T^{-1}_{g(X),s}(x)}\;\text{is increasing in}\;x,$$
% $$\frac{h\left(\ov T^{-1}_{X,s}(u)\right)}{g\left(\ov T^{-1}_{X,s}(u)\right)}\;\text{is increasing in}\;u\in[0,1].$$
%%%%%%%%%%%%%%%%%%%%%%%%%%%%%%%%%%%%%%%%%%%%%%%%%%%%%%%%%%%%%%%%%%%%%%%%%%%%%%%%%%%%%%%%%%%%%%%%%%%%%%%%%%%%%%%%%%%%%%%%%%%%%%%%%%%%%%%%%%%%%%%%%%%%%%%%%%%%%%%%%%%%%%%%%%%%%%%%%%%%%%%5
\section{s-NBU Ordering}
In this section we study s-NBU ordering.
\begin{d1}\label{def4-1} For any positive integer $s$, X (or its distribution $F_X$) is said to be more s-NBU than Y (or its distribution $F_Y$) (written as 
$F_X\leq_{s-NBU}F_Y$) if
$\alpha_s(x)\;\text{is super-additive}$.
\end{d1}\hfill $\Box$
%%%%%%%%%%%%%%%%%%%%%%%%%%%%%%%%%%%%%%%%%%%%%%%%%%%%%%%%%%%%%%%%%%%%%%%%%%%%%%%%%%%%%%%%%%%%%%%%%%%%%%%%%%%%%%%%%%%%%%%%%%%%%%%%%%%%%%%%%%%%%%%%%%%%%%%%%%%%%%%%%%%%%%%%%%%%%%%%%%%%%%%%%%%
%A function $f$ is called superadditive (resp. subadditive) if, for all $x,y$, $f(x+y)\geq(resp. \leq~)f(x)+f(y)$.
%%%%%%%%%%%%%%%%%%%%%%%%%%%%%%%%%%%%%%%%%%%%%%%%%%%%%%%%%%%%%%%%%%%%%%%%%%%%%%%%%%%%%%%%%%%%%%%%%%%%%%%%%%%%%%%%%%%%%%%%%%%%%%%%%%%%%%%%%%%%%%%%%%%%%%%%%%%%%%%%%%%%%%%%%%%%%%%%%%
\begin{r1}\label{rem4-1} For $s=1$, the above definition gives $F_X\leq_{NBU}F_Y$.
%or, equivalently, $F_Y^{-1}F_X(x)$ is super-additive.\hfill$\Box$
\end{r1}
%%%%%%%%%%%%%%%%%%%%%%%%%%%%%%%%%%%%%%%%%%%%%%%%%%%%%%%%%%%%%%%%%%%%%%%%%%%%%%%%%%%
%%%%%%%%%%%%%%%%%%%%%%%%%%%%%%%%%%%%%%%%%%%%%%%%%%%%
\begin{p1}
Definition \ref{def4-1} can equivalently be written as
 %\begin{enumerate}
   $$\ov T_{X,s}\left(\ov T^{-1}_{X,s}(u)+\ov T^{-1}_{X,s}(v)\right)\leq \ov T_{Y,s}\left(\ov T^{-1}_{Y,s}(u)+\ov T^{-1}_{Y,s}(v)\right)\;\text{for all}\;u,v \in [0,1].$$
 %\end{enumerate}
 \end{p1}
 {\bf Proof:} $F_X\leq_{s-NBU}F_Y$ holds if, and only if, for all $x,y\geq 0$,
 $$\overline{T}_{Y,s}^{-1}\left(\overline{T}_{X,s}(x+y)\right)\geq \overline{T}_{Y,s}^{-1}\left(\overline{T}_{X,s}(x)\right)+\overline{T}_{Y,s}^{-1}\left(\overline{T}_{X,s}(y)\right).$$
 Writing $x=\ov T^{-1}_{X,s}(u)$ and $y=\ov T^{-1}_{X,s}(v)$ in the above inequality, we get the required result.\hfill$\Box$
%%%%%%%%%%%%%%%%%%%%%%%%%%%%%%%%%%%%%%%%%%%%%%%%%%%%%%%%%%%%%%%%%%%%%%%%%%%%%%%%%%%5555   Remark %%%%%%%%%%%%%%%%%
%%%%%%%%%%%%%%%%%%%%%%%%%%%%%%%%%%%%%%%%%%%%%%%%%%%%%%%%%%%%%%%%%%%%%%%%%%%%%%%%%%%%%%%%%%%%%%%%%%%%%%%%%%%%%%%%%%%%%%%%%%%%%%%%%%%%%%%%%%%%%%%%%%%%%%%%%%%%%%%%%%%%%%%%%%%%%%%%%%%555
\\\hspace*{0.2 in} To prove the next theorem we use two lemmas which are given below. The proofs are omitted.
\begin{l1}\label{lem4-2}
 Let $f(\cdot)$ and $g(\cdot)$ be two nonnegative, increasing, and super-additive functions. Then $f\left(g(\cdot)\right)$ is super-additive.\hfill$\Box$
\end{l1}
% {\bf Proof:} Given that 
% \begin{eqnarray}\label{geq1}
% f(x+y)\geq f(x)+f(y)
% \end{eqnarray}
% and 
% \begin{eqnarray*}\label{geq2}
% g(x+y)\geq g(x)+g(y).
% \end{eqnarray*}
% Then 
% \begin{eqnarray*}
% f(g(x+y))&\geq& f(g(x)+g(y))
% \\&\geq& f(g(x))+f(g(y)),
% \end{eqnarray*}
% where the first inequality holds because $f(\cdot)$ is increasing function and the second inequality follows from (\ref{geq1}).
% Thus, $f\left(g(\cdot)\right)$ is super-additive.\hfill$\Box$
%%%%%%%%%%%%%%%%%%%%%%%%%%%%%%%%%%%%%%%%%%%%%%%%%%%%%%%%%%%%%%%%%%%%%%%%%%%%%%%%%%%%%%%%%%%%%%%%%%%%%%%%%%%%%%%%%%%%%%%%%%%%%%%%%%%%%%%%%%%%%%%%%%%%%%%%%%%%%%%%%%%%%%%%%%%%%%%%%%%%%%%
\begin{l1}\label{lem4-1}
 Let $f(\cdot)$ be a nonnegative, increasing, and super-additive function. Then $f^{-1}(\cdot)$ is sub-additive.\hfill$\Box$
\end{l1}
% {\bf Proof:} Given that $f(x+y)\geq f(x)+f(y)$. Write $f(x)=\alpha$ and $f(y)=\beta$. Then 
% \begin{eqnarray*}
%  f^{-1}(\alpha+\beta)&=&f^{-1}(f(x)+f(y)\\
%  &\leq& f^{-1}(f(x+y))\\
%  &=&x+y\\
%  &=&f^{-1}(\alpha)+f^{-1}(\beta).
%  \end{eqnarray*}
%  Thus, $f^{-1}(\cdot)$ is sub-additive.\hfill$\Box$
 %%%%%%%%%%%%%%%%%%%%%%%%%%%%%%%%%%%%%%%%%%%%%%%%%%%%%%%%%%%%%%%%%%%%%%%%%%%%%%%%%%%%%%%%%%%%%%%%%%%%%%%%%%%%%%%%%%%%%%%%%%%%%%%%%%%%%%%%%%%%%%%%%%%%%%%%%%5555%%%%%%%%%%%%%%%%%%%%%%%%%%%
\hspace*{0.2 in}The following theorem shows that s-NBU ordering is a partial ordering.
 \begin{t1}\label{th4-1} The relationship $F_X\leq_{s-NBU}F_Y$ is a partial ordering of the equivalence classes of $\cal{F}$.
\end{t1}
{\bf Proof:} ($i$) The proof of reflexive property of s-NBU ordering is trivial.
\\($ii$) Let $F_X\leq_{s-NBU}F_Y$. Then 
$${T}_{Y,s}^{-1}\left({T}_{X,s}(x)\right)\;\text{is super-additive}.$$
By Lemma \ref{lem4-1}, the above statement can equivalently be written as 
\begin{equation}
{T}_{X,s}^{-1}\left({T}_{Y,s}(x)\right)\;\text{is sub-additive}.\label{eqn4-1}
\end{equation}
Further,
$F_Y\leq_{s-NBU}F_X$ gives that
\begin{equation}
{T}_{X,s}^{-1}\left({T}_{Y,s}(x)\right)\;\text{is super-additive}.\label{eqn4-2}
\end{equation}
Combining ($\ref{eqn4-1}$) and ($\ref{eqn4-2}$), we get 
$${T}_{X,s}^{-1}\left({T}_{Y,s}(x)\right)=\beta x,$$
 for some constant $\beta$,
which, by Lemma \ref{lem2-4}, gives $F_X\sim F_Y$.\\
($iii$) On using Lemma \ref{lem4-2}, one can easily verify that s-NBU ordering is transitive.\hfill$\Box$
%%%%%%%%%%%%%%%%%%%%%%%%%%%%%%%%%%%%%%%%%%%%%%%%%%%%%%%%%%%%%%%%%%%%%%%%%%%%%%%%%%%%%%%%%%%%%%%%%%%%%%%%%%%%%%%%%%%%%%%%%%%%%%%%%%%%%%%%%%%%%%%%%%%%%%%%%%%%%%%%%%%%%%%%%%%%%%%%%%%%%%%%%
\\\hspace*{0.2 in}Below Theorem \ref{th4-2} shows that, if a probability distribution is smaller than exponential distribution in s-NBU ordering, then it is actually an s-NBU distribution. 
 The proof follows from Lemma \ref{lem2-3}.
\begin{t1}\label{th4-2} Let $\overline{F}_Y(x)=e^{-\lambda x},\; \lambda >0$. Then, for $s=1,2,\ldots$,
$$F_X\leq_{s-NBU}F_Y\;\text{if, and only if,}\; F_X\;\text{is s-NBU}.$$
\end{t1}\hfill$\Box$
%%%%%%%%%%%%%%%%%%%%%%%%%%%%%%%%%%%%%%%%%%%%%%%%%%%%%%%%%%%%%%%%%%%%%%%%%%%%%%%%%%%%%%%%%%%%%%%%
\\\hspace*{0.2 in}Since, all star-shaped functions are super-additive, we have the following theorem.
\begin{t1}\label{rem4-2} 
If $F_X\leq_{s-IFRA}F_Y$, then $F_X\leq_{s-NBU}F_Y.$\hfill$\Box$
\end{t1}
%%%%%%%%%%%%%%%%%%%%%%%%%%%%%%%%%%%%%%%%%%%%%%%%%%%%%%%%%%%%%%%%%%%%%%%%%%%%%%%%%%%%%%%%%%%%%%%%%%%%%%%%%%%%%%%%%%%%%%%%%%%%%%%%%%%%%%%%%%%%%%%%%%%%%%%%%%%%%%%%%%%%%%%%%%%%%%559999999999999
                                                     %%%%%%%%%%%%%%%%%%  s-NBUFR   %%%%%%%%%%%%%%%%%%%%%%%%%%%%%%%
%%%%%%%%%%%%%%%%%%%%%%%%%%%%%%%%%%%%%%%%%%%%%%%%%%%%%%%%%%%%%%%%%%%%%%%%%%%%%%%%%%%%%%%%%%%%%%%%%%%%%%%%%%%%%%%%%%%%%%%%%%%%%%%%%%%%%%%%%%%%%%%%%%%%%%%%%%%%%%%%%%%%%%%%%%%%%%%%%%%%%%%%%55555                                                     
\section{s-NBUFR Ordering}\label{sec5-1}
We begin this section with the following definition.
\begin{d1}\label{def5-1} For any positive integer $s$,
 X (or its distribution $F_X$) is said to be more s-NBUFR than Y (or its distribution $F_Y$) (written as $F_X\leq_{s-NBUFR}F_Y$)
if $\alpha'_s(x)\geq \alpha'_s(0).$~\hfill$\Box$
\end{d1}
%%%%%%%%%%%%%%%%%%%%%%%%%%%%%%%%%%%%%%%%%%%%%%%%%%%%%%%%%%%%%%%%%%%%%%%%%%%%%%%%%%%%%%%%%%%%%%%%%%%%%%%%%%%%%%%%%%%%%%%%%%%%%%%%%%%%%%%%%%%%%%%%%%%%%%%%%%%%%%%%%%%%%%%%%%%%%%%%%%%
\begin{r1}\label{rem5-1} For $s=1$, $s=2$ and $s=3$, the above definition gives $F_X\leq_{NBUFR}F_Y$,
$F_X\leq_{NBUE}F_Y$ and $F_X\leq_{NDVRL}F_Y$, respectively.
%NBUFR and NBUE orderings are studied by Kochar and Wiens~\cite{kw1}.\hfill$\Box$
\end{r1}
%%%%%%%%%%%%%%%%%%%%%%%%%%%%%%%%%%%%%%%%%%%%%%%%%%%%%%%%%%%%%%%%%%%%%%%%%%%%%%%%%%%%%%%%%%%%%%%%%%%%%%%%%%%%%%%%%%%%%%%%%%%%%%%%%%%%%%%%%%%%%%%%%%%%%%%%%%%%%%%%%%%%%%%%%%%%%%%%%%%%%
\hspace*{0.2 in}In the following proposition we discuss some equivalent conditions of the s-NBUFR ordering.
\begin{p1}\label{pp4}
 For $s=2,3,\dots,$ Definition \ref{def5-1} can equivalently be written in one of the following forms:
 \begin{enumerate}
 \item [$(i)$] $\alpha_s(x)\geq \alpha_{s-1}(x)$ for all $x\geq 0$.
 \item [$(ii)$] $\frac{r_{X,s}\left(T^{-1}_{X,s}(u)\right)}{r_{Y,s}\left(T^{-1}_{Y,s}(u)\right)}\geq \frac{\wt \mu_{Y,s-1}}{\wt \mu_{X,s-1}}$ for all $u\in[0,1]$.
 \item [$(iii)$] $\frac{r_{X,s}\left(T^{-1}_{X,s-1}(u)\right)}{r_{Y,s}\left(T^{-1}_{Y,s-1}(u)\right)}\geq \frac{\wt \mu_{Y,s-1}}{\wt \mu_{X,s-1}}$ for all $u\in[0,1]$.
  \item [$(iv)$] $\frac{\mu_{Y,s-1}\left(T^{-1}_{Y,s}(u)\right)}{\mu_{X,s-1}\left(T^{-1}_{X,s}(u)\right)}\geq \frac{\wt \mu_{Y,s-1}}{\wt \mu_{X,s-1}}$ for all $u\in[0,1]$.
  \item [$(v)$] $\frac{\mu_{Y,s-1}\left(T^{-1}_{Y,s-1}(u)\right)}{\mu_{X,s-1}\left(T^{-1}_{X,s-1}(u)\right)}\geq \frac{\wt \mu_{Y,s-1}}{\wt \mu_{X,s-1}}$ for all $u\in[0,1]$.
  \item [$(vi)$] $\frac{\ov T_{Y,s-1}\left(\alpha_s(x)\right)}{\ov T_{Y,s-1}\left(\alpha_{s-1}(x)\right)}\leq 1$ for all $x \geq 0$.
  \item [$(vii)$] $\mathcal{R}_{X,s-1}(u)\geq \mathcal{R}_{Y,s-1}(u)\;\text{for all} \;u\in [0,1]$.
  \item [$(viii)$] $\mathcal{H}_{X,s-1}(u)\leq \mathcal{H}_{Y,s-1}(u)\;\text{for all} \;u\in [0,1]$.
 \end{enumerate}
 \end{p1}
 {\bf Proof:} $F_X\leq_{NBUFR}F_Y$ if, and only if, for all $x\geq 0$,
 \begin{eqnarray}\label{mre55}
 \ov T_{X,s-1}(x)\geq \ov T_{Y,s-1}\ov T^{-1}_{Y,s}\left(\ov T_{X,s}(x)\right),
 \end{eqnarray}
%  Clearly, for $s=1$, 
%  $$f(x)\leq \frac{f(0)}{g(0)}g\bar G^{-1}\bar F(x).$$
%  For $s=2,3,\dots,$ $(\ref{mre55})$ becomes
%The above inequality holds if, and only if,
 %$$\ov T_{X,s-1}(x)\geq \ov T_{Y,s-1}\ov T^{-1}_{Y,s}\left(\ov T_{X,s}(x)\right),$$
 or equivalently,
 $$\alpha_s(x)\geq \alpha_{s-1}(x),$$ which is $(i)$. 
 %%%%%%%%%%%%%%%%%%%%%%%%%  Proof b  555555555555555555
  Note that, for all $x\geq 0$, (\ref{mre55}) can equivalently be written as
 $$\frac{r_{X,s}(x)}{r_{Y,s}\left(\ov T^{-1}_{Y,s}\ov T_{X,s}(x)\right)}\geq \frac{\wt \mu_{Y,s-1}}{\wt \mu_{X,s-1}},$$
 or equivalently,
$$\frac{r_{X,s}\left(T^{-1}_{X,s}(u)\right)}{r_{Y,s}\left(T^{-1}_{Y,s}(u)\right)}\geq \frac{\wt \mu_{Y,s-1}}{\wt \mu_{X,s-1}}\quad\text{for all}\;u\in[0,1].$$
This proves the equivalence of $(i)$ and $(ii)$.
%%%%%%%%%%%%%%%%%%%%%%%%%%%%%%%%%   Proof c 555555555555555555
Now, for all $u\in [0,1],$
$$\frac{r_{X,s}\left(T^{-1}_{X,s-1}(u)\right)}{r_{Y,s}\left(T^{-1}_{Y,s-1}(u)\right)}\geq \frac{\wt\mu_{Y,s-1}}{\wt\mu_{X,s-1}}$$
holds if, and only if, 
$$\frac{r_{X,s}(x)}{r_{Y,s}\left(\ov T^{-1}_{Y,s-1}\ov T_{X,s-1}(x)\right)}\geq \frac{\wt\mu_{Y,s-1}}{\wt \mu_{X,s-1}}\quad\text{for all}\;x\geq0.$$
The above inequality can equivalently be written as
$$\ov T_{Y,s}\left(\ov T^{-1}_{Y,s-1}\ov T_{X,s-1}(x)\right)\geq \ov T_{X,s}(x),$$
or equivalently,
$$\alpha_s(x)\geq \alpha_{s-1}(x)\quad\text{for all}\;x\geq0,$$ giving the equivalence of $(i)$ and $(iii)$.
 On using (\ref{mre5}) in $(ii)$ and $(iii)$, we get $(iv)$ and $(v)$, respectively. 
Equivalence of $(i)$, and $(vi)$ and $(vii)$ follows
 from (\ref{mre55}).
%%%%%%%%%%%%%%%%%  Proof g & h%%%%%%%%%%%%%%%%%%%%%%%%%%
Equivalence of $(vii)$ and $(viii)$ follows from (\ref{mre76}).
\hfill $\Box$ 
%%%%%%%%%%%%%%%%%%%%%%%%%%%%%%%%%%%%%%%%%%%%%%%%%%%%%%%%%%%%%%%%%%%%%%%%%%%%%%%%%%%%%%%%%%%%%   Theorem  5.1  %%%%%%%%%%%%%%%%%%%%%%%%%%%%%%%%%%%%%%%%%%%%%%%%%%%%%%%%%%%%%%%%%%%%%%%
\\\hspace*{0.2 in}The following theorem shows that s-NBUFR ordering is a partial ordering. 
\begin{t1} \label{th5-1}
The relationship $F_X\leq_{s-NBUFR}F_Y$ is a partial ordering of the equivalence
classes of $\cal{F}$.
\end{t1}
{\bf Proof:} For $s=1$, the result follows from Kochar and Wiens~\cite{kw1}. We only prove the result for $s=2,3,\dots$ $\cdot$ Let us fix $s$.
\\($i$) It is easy to verify that s-NBUFR ordering is reflexive.\\
($ii$) By Proposition \ref{pp4}(i), $F_X\leq_{s-NBUFR}F_Y$ holds if, and only if,
\begin{eqnarray}\label{mre56}
 \overline{T}_{Y,s-1}^{-1}(\overline{T}_{X,s-1}(x))\leq \overline{T}_{Y,s}^{-1}(\overline{T}_{X,s}(x)).
\end{eqnarray}
Further, $F_Y\leq_{s-NBUFR}F_X$ gives 
\begin{eqnarray}\label{mre57}
 \overline{T}_{X,s-1}^{-1}(\overline{T}_{Y,s-1}(x))\leq \overline{T}_{X,s}^{-1}(\overline{T}_{Y,s}(x)).
\end{eqnarray}
Replacing $x$ by $\overline{T}_{Y,s}^{-1}(\overline{T}_{X,s}(x))$ in (\ref{mre57}), we have
\begin{eqnarray}\label{mre58}
 \overline{T}_{Y,s}^{-1}(\overline{T}_{X,s}(x))\leq \overline{T}_{Y,s-1}^{-1}(\overline{T}_{X,s-1}(x)).
\end{eqnarray}
Combining ($\ref{mre56}$) and ($\ref{mre58}$), we get
\begin{eqnarray}\label{mre59}
\alpha_{s}(x)=\alpha_{s-1}\; \text{for all}\;x\geq 0. 
\end{eqnarray}
Note that
\begin{eqnarray}
 \alpha'_s(x)&=&\left(\frac{\wt\mu_{Y,s-1}}{\wt\mu_{X,s-1}}\right)\left(\frac{\ov T_{Y,s-1}\left(\alpha_{s-1}(x)\right)}{\ov T_{Y,s-1}\left(\alpha_s(x)\right)}\right)\nonumber
\\&=&\theta,\label{mre60}
\end{eqnarray}
where the last equality follows from (\ref{mre59}) and $\theta=\wt\mu_{Y,s-1}/\wt\mu_{X,s-1}$ (constant). Now, integrating (\ref{mre60}) from $0$ to $x$,
and then using $\alpha_s(0)=0$,
we have $\ov T_{X,s}(x)=\ov T_{Y,s}(\theta x).$
Thus, on using Lemma \ref{lem2-4}, we have $F_X\sim F_Y$.
\\($iii$) $F_X\leq_{s-NBUFR}F_Y$ gives 
\begin{equation}\label{mre61}
\overline{T}_{Y,s-1}^{-1}(\overline{T}_{X,s-1}(x))\leq \overline{T}_{Y,s}^{-1}(\overline{T}_{X,s}(x))
\end{equation}
and $F_Y\leq_{s-NBUFR}F_Z$ gives 
\begin{equation}\label{mre62}
\overline{T}_{Z,s-1}^{-1}(\overline{T}_{Y,s-1}(x))\leq \overline{T}_{Z,s}^{-1}(\overline{T}_{Y,s}(x)).
\end{equation}
Now, 
\begin{eqnarray*}
 \overline{T}_{Z,s-1}^{-1}(\overline{T}_{X,s-1}(x))&=&\overline{T}_{Z,s-1}^{-1}\overline{T}_{Y,s-1}\left(\overline{T}_{Y,s-1}^{-1}\overline{T}_{X,s-1}(x)\right)
 \\&\leq&\overline{T}_{Z,s-1}^{-1}\overline{T}_{Y,s-1}\left(\overline{T}_{Y,s}^{-1}\overline{T}_{X,s}(x)\right)
 \\&\leq&\overline{T}_{Z,s}^{-1}\overline{T}_{Y,s}\left(\overline{T}_{Y,s}^{-1}\overline{T}_{X,s}(x)\right)
 \\&=&\overline{T}_{Z,s}^{-1}(\overline{T}_{X,s}(x)),
\end{eqnarray*}
where the first inequality follows from (\ref{mre61}) and using the fact that $\overline{T}_{Z,s-1}^{-1}\overline{T}_{Y,s-1}(\cdot)$ is an increasing function. The second inequality
holds from (\ref{mre62}). Thus, s-NBUFR ordering is transitive.~\hfill $\Box$
%%%%%%%%%%%%%%%%%%%%%%%%%%%%%%%%%%%%%%%%%%%%%%%%%%%%%%%%%%%%%%%%%%%%%%%%%%%%%%%%%%%%%%%%%%%%%%%%%%%%%%%%%%%%%%%%%%%%%%%%%%%%%%%%%%%%%%%%%%%%%%%%%%%%%%%%%%%%%%%%%%%%%%%%%%%%%%%%%%%%%%%%%5
\\\hspace*{0.2 in}The following theorem shows that a random variable $X$ is s-NBUFR if, and only if, $X$ is smaller than exponential distribution in s-NBUFR ordering.
The proof follows from Lemma~\ref{lem2-3}.
\begin{t1}
If $\overline{F}_Y(x)=e^{-\lambda x},\; \lambda>0$, then 
$F_X\leq_{s-NBUFR}F_Y$ if, and only if, $F_X$ is s-NBUFR.~\hfill$ \Box$
\end{t1}
%%%%%%%%%%%%%%%%%%%%%%%%%%%%%%%%%%%%%%%%%%%%%%%%%%%%%%%%%%%%%%%%%%%%%%%%%%%%%%%%%%%%%%%%%%%%%%%%%%%%%%%%%%%%%%%%%%%%%%%%%%%%%%%%%%%%%%%%%%%%%%%%%%%%%%%%%%%%%%%%%%%%%%%%%%%%%%%%%
\hspace*{0.2 in}In the following theorem, we prove that s-NBU ordering implies s-NBUFR ordering. 
\begin{t1}\label{th5-4}
$F_X\leq_{s-NBU}F_Y\;\Rightarrow F_X\leq_{s-NBUFR}F_Y$.
\end{t1}
{\bf Proof:} $F_X\leq_{s-NBU}F_Y$ gives that, for all $x,y\geq 0$,
$$\alpha_s(x+y)\geq \alpha_s(x)+\alpha_s(y).$$
Taking limit as $y\to 0$ on both sides of the above inequality, and then using $\alpha_s(0)=0$, we get the required result.\hfill $\Box$
%%%%%%%%%%%%%%%%%%%%%%%%%%%%%%%%%%%%%%%%%%%%%%%%%%%%%%%%%%%%%%%%%%%%%%%%%%%%%%%%%%%%%%%%%%%%%%%%%%%%%%%%%%%%%%%%%%%%%%%%%%%%%%%%%%%%%%%%%%%%%%%%%%%%%%%%%%%%%%%%%%%%%%%%%%%5000000000000000000
                                                              %%%%%%%%%%%%%   s-NBUAFR    %%%%%%%%%%%%%%%%%%%%%%%%%%%%
%%%%%%%%%%%%%%%%%%%%%%%%%%%%%%%%%%%%%%%%%%%%%%%%%%%%%%%%%%%%%%%%%%%%%%%%%%%%%%%%%%%%%%%%%%%%%%%%%%%%%%%%%%%%%%%%%%%%%%%%%%%%%%%%%%%%%%%%%%%%%%%%%%%%%%%%%%%%%%%%%%%%%%%%%%%%%%%%%%%%%%%%%%%
 \section{s-NBAFR Ordering}\label{sec6-1}
 In this section we study s-NBAFR ordering. We start with the following definition. 
\begin{d1}\label{def6-1}For any positive integer $s$,
 X (or its distribution $F_X$) is said to be more s-NBAFR than Y (or its distribution $F_Y$) (written as $F_X\leq_{s-NBAFR} F_Y$) if
$\alpha_s(x)\geq x\alpha'_s(0).$~\hfill $\Box$
\end{d1}
%%%%%%%%%%%%%%%%%%%%%%%%%%%%%%%%%%%%%%%%%%%%%%%%%%%%%%%%%%%%%%%%%%%%%5%%%%%%%%%%%%%%%%%%%%%%%%%%%%%%%%%%%%%%%%%%%%%%%%%%%%%%%%%%%%%%%%%%%%%%%%%%%%%%%%%%%%%%%%%%%%%%%%%%%%%%%%%%%%%%%
%%%%%%%%%%%%%%%%%%%%%%%%%%%%%%%%%%%%%%%%%%%%%%%%%%%%%%%%%%%%%  Remark  %%%%%%%%%%%%%%%%%%%%%%%%%%%%%%%%%%%%%%%%%%%
\begin{r1}\label{rem6-1} For $s=1$ and $s=2$, the above definition gives $F_X\leq_{NBAFR}F_Y$ and
$F_X\leq_{HNBUE}F_Y$, respectively. 
%These two orderings are discussed in Kochar and Wiens~\cite{kw1}. 
\end{r1}\hfill $\Box$
%%%%%%%%%%%%%%%%%%%%%%%%%%%%%%%%%%%%%%%%%%%%%%%%%%%%%%%%%%%%%%%%%%%%%%%%%%%%%%%%%%%%%%%%%%%%%%%%%%%%%%%%%%%%%%%%%%%%%%%%%%%%%%%%%%%%%%%%%%%%%%%%%%%%%%%%%%%%%%%%%%%%%%%%
\\\hspace*{0.2 in}Below we give some equivalent representations of the s-NBAFR ordering. 
\begin{p1}\label{pp6}
  For $s=2,3,\dots$, Definition \ref{def6-1} can equivalently be written in one of the following forms:
  \begin{enumerate}
  \item [$(i)$] $\ov T_{X,s}\left(x\wt \mu_{X,s-1}\right)\leq \ov T_{Y,s}\left(x\wt\mu_{Y,s-1}\right)\;\text {for all}\;x\geq 0.$
  \item [$(ii)$] $\mathcal{H}^{-1}_{X,s-1}(u)\geq \mathcal{H}^{-1}_{Y,s-1}\left[T_{Y,s-1}\left(\frac{\wt \mu_{Y,s-1}}{\wt \mu_{X,s-1}}T^{-1}_{X,s-1}(u)\right)\right]\;\text{for all} \;u\in [0,1]$.
   \item [$(iii)$] $\mathcal{R}^{-1}_{X,s-1}(u)\leq \mathcal{R}^{-1}_{Y,s-1}\left[T_{Y,s-1}\left(\frac{\wt \mu_{Y,s-1}}{\wt \mu_{X,s-1}}T^{-1}_{X,s-1}(u)\right)\right]\;\text{for all} \;u\in [0,1]$.
  \item [$(iv)$] $\mathcal{L}_{X,s-1}(u)\geq \mathcal{L}_{Y,s-1}(u)\;\text{for all} \;u\in[0,1].$
  \end{enumerate}
 \end{p1}
 {\bf Proof:} $F_X\leq_{s-NBAFR}F_Y$ holds if, and only, if, for all $x\geq 0$,
 \begin{eqnarray}
 \overline{T}_{Y,s}^{-1}(\overline{T}_{X,s}(x))\geq \frac{\wt \mu_{Y,s-1}}{\wt \mu_{X,s-1}}x.\label{mre71}
 \end{eqnarray}
 Replacing $x$ by $x\wt \mu_{X,s-1}$ in (\ref{mre71}), we get $(i)$.
Note that (\ref{mre71}) holds if, and only if, for all $x\geq 0$,
\begin{eqnarray}\label{mre72}
{T}_{X,s}(x)\geq {T}_{Y,s}\left(\frac{\wt \mu_{Y,s-1}}{\wt \mu_{X,s-1}}x\right).
\end{eqnarray}
Putting $x=T^{-1}_{X,s-1}(u)$ in (\ref{mre72}), we see that $(i)$ and $(ii)$ are equivalent.
On using $(\ref{mre70})$, $(ii)$ and $(iii)$ become equivalent. 
Now, $(i)$ can equivalently be written as 
$$\int\limits_x^{\infty}\ov T_{X,s-1}\left(t\wt \mu_{X,s-1}\right)dt\leq \int\limits_x^{\infty}\ov T_{Y,s-1}\left(t\wt \mu_{Y,s-1}\right)dt,$$
or equivalently,
\begin{eqnarray}
\int\limits_x^{\infty}\ov T_{X^*,s-1}\left(t\right)dt\leq \int\limits_x^{\infty}\ov T_{Y^*,s-1}\left(t\right)dt,\label{mre77}
\end{eqnarray}
where $\ov T_{X^*,s-1}(t)=\ov T_{X,s-1}\left(t\wt \mu_{X,s-1}\right)$ and $\ov T_{Y^*,s-1}(t)=\ov T_{Y,s-1}\left(t\wt \mu_{Y,s-1}\right)$
be the respective survivals of two random variables $X^*$ and $Y^*$. 
%Then, (\ref{mre77}) represents
%$$X^*\leq_{st2}Y^*.$$
Thus, on using Theorem 4 of Taillie~\cite{t6}, $(\ref{mre77})$ can equivalently be written as
$$\frac{1}{\wt \mu_{X,s-1}}\int\limits_0^uT^{-1}_{X,s-1}(t)dt\geq \frac{1}{\wt \mu_{Y,s-1}}\int\limits_0^uT^{-1}_{Y,s-1}(t)dt\quad\text{for all}\;u \in [0,1].$$
This proves the equivalence of $(i)$ and $(iv)$.\hfill $\Box$
\\\hspace*{0.2 in}The following theorem shows that s-NBAFR ordering is a partial ordering.
%%%%%%%%%%%%%%%%%%%%%%%%%%%%%%%%%%%%%%%%%%%%%%%%%%%%%%%%%%%%%%%%%%%%%%%%%%%%%%%%%%%%%%%%%%%%%   Theorem  6.1  %%%%%%%%%%%%%%%%%%%%%%%%%%%%%%%%%%%%%%%%%%%%%%%%%%%%%%%%%%%%%%%%%%%%%%%
\begin{t1} The relationship $F_X\leq_{s-NBAFR}F_Y$ is a partial ordering of the equivalence
classes of $\cal{F}$.
\end{t1}
{\bf Proof:} For $s=1$, the result follows from Kochar and Wiens~\cite{kw1}. We only prove the result for $s=2,3,\dots$ $\cdot$ Let us fix $s$.
\\($i$) That s-NBAFR ordering is reflexive, is trivial.\\
($ii$) $F_X\leq_{s-NBAFR}F_Y$ gives 
\begin{eqnarray}\label{mre63}
 \overline{T}_{Y,s}^{-1}(\overline{T}_{X,s}(x))\geq \frac{\wt \mu_{Y,s-1}}{\wt \mu_{X,s-1}}x
\end{eqnarray}
and $F_Y\leq_{s-NBAFR}F_X$ gives 
\begin{eqnarray}\label{mre64}
 \overline{T}_{X,s}^{-1}(\overline{T}_{Y,s}(x))\geq \frac{\wt \mu_{X,s-1}}{\wt \mu_{Y,s-1}}x.
\end{eqnarray}
Replacing $x$ by $\overline{T}_{Y,s}^{-1}(\overline{T}_{X,s}(x))$ in (\ref{mre64}), we have
\begin{eqnarray}\label{mre65}
 \overline{T}_{Y,s}^{-1}(\overline{T}_{X,s}(x))\leq \frac{\wt \mu_{Y,s-1}}{\wt \mu_{X,s-1}}x.
\end{eqnarray}
Combining (\ref{mre63}) and (\ref{mre65}), we have
\begin{eqnarray*}
 \overline{T}_{Y,s}^{-1}(\overline{T}_{X,s}(x))&=& \frac{\wt \mu_{Y,s-1}}{\wt \mu_{X,s-1}}x
 \\&=&\theta x,
\end{eqnarray*}
where $\theta=\wt \mu_{Y,s-1}/\wt \mu_{X,s-1}\;$(constant). Hence, $\ov T_{X,s}(x)=\ov T_{Y,s}(\theta x).$
Thus, on using Lemma \ref{lem2-4}, we have $F_X\sim F_Y$.
\\($iii$)  $F_X\leq_{s-NBAFR}F_Y$ gives
\begin{equation}\label{mre66}
\overline{T}_{Y,s}^{-1}(\overline{T}_{X,s}(x))\geq \frac{\wt \mu_{Y,s-1}}{\wt \mu_{X,s-1}}x
\end{equation}
and $F_Y\leq_{s-NBAFR}F_Z$ gives
\begin{equation}\label{mre67}
\overline{T}_{Z,s}^{-1}(\overline{T}_{Y,s}(x))\geq \frac{\wt \mu_{Z,s-1}}{\wt \mu_{Y,s-1}}x.
\end{equation}
Now, 
\begin{eqnarray*}
 \overline{T}_{Z,s}^{-1}(\overline{T}_{X,s}(x))&=&\overline{T}_{Z,s}^{-1}\overline{T}_{Y,s}\left(\overline{T}_{Y,s}^{-1}\overline{T}_{X,s}(x)\right)
 \\&\geq&\overline{T}_{Z,s}^{-1}\overline{T}_{Y,s}\left(\frac{\wt \mu_{Y,s-1}}{\wt \mu_{X,s-1}}x\right)
 \\&\geq&\left(\frac{\wt \mu_{Z,s-1}}{\wt \mu_{Y,s-1}}\right)\left(\frac{\wt \mu_{Y,s-1}}{\wt \mu_{X,s-1}}x\right)
 \\&=&\frac{\wt \mu_{Z,s-1}}{\wt \mu_{X,s-1}}x,
\end{eqnarray*}
where the first inequality holds from (\ref{mre66}) and using the fact that $\overline{T}_{Z,s}^{-1}\overline{T}_{Y,s}(\cdot)$ is an increasing function. The second inequality
holds from (\ref{mre67}). Thus, s-NBAFR ordering is transitive.\hfill $\Box$
%%%%%%%%%%%%%%%%%%%%%%%%%%%%%%%%%%%%%%%%%%%%%%%%%%%%%%%%%%%%%%%%%%%%%%%%%%%%%%  Theorem 6.2 %%%%%%%%%%%%%%%%%%%%%%%%%%%%%%%%%
\\\hspace*{0.2 in}In the following theorem we represent the relationship between s-NBAFR ageing and s-NBAFR ordering. The proof follows from Lemma~\ref{lem2-3}
\begin{t1}\label{th6-4}
If $\overline{F}_Y(x)=e^{-\lambda x},\; \lambda>0$, then 
$F_X\leq_{s-NBAFR}F_Y$ if, and only if, $F_X$ is s-NBAFR.~\hfill$ \Box$
\end{t1}
%%%%%%%%%%%%%%%%%%%%%%%%%%%%%%%%%%%%%%%%%%%%%%%%%%%%%%%%%%%%%%%%%%%%%%%%%%%%%%%%%%%%%% Theorem 6.3  %%%%%%%%%%%%%%%%%%%%%%%%%%%%%
\hspace*{0.2 in}Below we show that s-NBUFR ordering implies s-NBAFR ordering.
\begin{t1}\label{th6-5}
$F_X\leq_{s-NBUFR}F_Y\;\Rightarrow F_X\leq_{s-NBAFR}F_Y$.
\end{t1}
{\bf Proof:} $F_X\leq_{s-NBUFR}F_Y$ gives that, for all $x\geq0$,
$$\alpha'_s(x)\geq \alpha'_s(0).$$
Integrating with limit from $0$ to $x$ on both sides of the above inequality, and then using $\alpha_s(0)=0$, we get the required result.\hfill $\Box$
%%%%%%%%%%%%%%%%%%%%%%%%%%%%%%%%%%%%%%%%%%%%%%%%%%%%%%%%%%%%%%%%%%%%%%%%%%%%%%%%%%%%%%%%%%%%%%%%%%%%%%%%%%%%%%%%%%%%%%%%%%%%%%%%%%%%%%%%%%%%%%%%%%%%%%%%%%%%%%%%%%%%%%%%%%%%%%%%%%%%%%%%%%%%
                                                    %%%%%%%%%%%%%%%%%%%%%%%%%%%%%%  Conclusion %%%%%%%%%%%%%%%%%%%%%%%%%%%%%%%%%%%%%%%%
%%%%%%%%%%%%%%%%%%%%%%%%%%%%%%%%%%%%%%%%%%%%%%%%%%%%%%%%%%%%%%%%%%%%%%%%%%%%%%%%%%%%%%%%%%%%%%%%%%%%%%%%%%%%%%%%%%%%%%%%%%%%%%%%%%%%%%%%%%%%%%%%%%%%%%%%%%%%%%%%%%%%%%%%%%%%%%%%%%%%%%%%%%%55555
\section{Concluding Remarks}\label{sec7-1}
\hspace*{0.3 in}In this paper we introduce some new generalized partial orderings. We give some equivalent representations of each generalized ordering in 
terms of failure rate function, mean residual life function, TTT transform, Lorenz curve, etc.
We discuss an alternative way out to study the generalized ageings in terms of generalized orderings.
These orderings throw new light on the understanding of the phenomenon of generalized ageings. 
Such a study is meaningful because it summarizes the existing results available in literature in a unified way.
Further, the lives of two systems may have same ageing property, but one may age faster than the other. So, one might be interested to know which one is ageing slower to decide on which 
of the two systems to be chosen. The ageing orderings help one to decide on this. Again, if one group of components are known to have the less rate of ageing compared to the other set, 
this will help the design engineers to select the former group of components in place of the latter group while designing a system.
We conclude our discussion by mentioning the following
chain of implications of generalized orderings.
\\\hspace*{1 in}$F_X\leq_{s-IFR}F_Y\Rightarrow F_X\leq_{s-IFRA}F_Y$
\\\hspace*{2.5 in}$\Downarrow$
\\\hspace*{2.2 in}~$F_X\leq_{s-NBU}F_Y$
\\\hspace*{2.5 in}$\Downarrow$
\\\hspace*{2.2 in}~$ F_X\leq_{s-NBUFR}F_Y\Rightarrow F_X\leq_{s-NBAFR}F_Y.$
%%%%%%%%%%%%%%%%%%%%%%%%%%%%%%%%%%%%%%%%%%%%%%%%%%%%%%%%%%  Acknowledgements  %%%%%%%%%%%%%%%%%%%%%%%%%%%%%%%%%%%%%%
\section*{Acknowledgements}
%\hspace*{0.3 in}The authors gratefully acknowledge the constructive comments of the referee which lead to an improved version of the manuscript.
  Financial support from Council of Scientific and Industrial Research, New Delhi (CSIR Grant No. 09/921(0060)2011-EMR-I) is sincerely
 acknowledged by Nil Kamal Hazra. 
%%%%%%%%%%%%%%%%%%%%%%%%%%%%%%%%%%%%%%%%%%%%%%%%%%%%%%%%%%%%%%%%%%%%%%%%%%%%%%%%%%%%%%%%%%%%%%%%%%%%%%%%%%%%%%%%%%%%%%%%%%%%%%%%%%%%%%%%%%%%%%%%%%%%%%%%%%%%%%%%%%%%%%%%%%%%%%%%%%%%%%555555555555555555
                                                              %%%%%%%%%%%%%%%%%%   References  %%%%%%%%%%%%%%%%%%%%%%%%%%%
%%%%%%%%%%%%%%%%%%%%%%%%%%%%%%%%%%%%%%%%%%%%%%%%%%%%%%%%%%%%%%%%%%%%%%%%%%%%%%%%%%%%%%%%%%%%%%%%%%%%%%%%%%%%%%%%%%%%%%%%%%%%%%%%%%%%%%%%%%%%%%%%%%%%%%%%%%%%%%%%%%%%%%%%%%%%%%%%%%%%%%%%%%%%%%                                                              
%\section*{References}

\end{document}